\documentclass[twocolumn,showpacs,preprintnumbers,amsmath,amssymb,aps,prb,superscriptaddress]{revtex4-2}

\pdfoutput=1
\usepackage[pdftex]{graphicx}
\usepackage[english]{babel}
\usepackage{soul}
\usepackage{cleveref}
\usepackage{bbold}
\usepackage{mathtools}
\usepackage{multirow}
\usepackage{colortbl}
\definecolor{kugray5}{RGB}{224,224,224}

\usepackage{color}
\usepackage{amstext}
\usepackage{braket}
\usepackage{mathdots}
\usepackage{tikz}
\usepackage{physics}
\usepackage{enumitem}
\usepackage{diagbox}
\usepackage{mciteplus}

\usepackage{dcolumn}
\usepackage{bm}

\usetikzlibrary{matrix,decorations.pathreplacing}

\begin{document}


\title{$2^n$-root weak, Chern, and higher-order topological insulators, and $2^n$-root topological semimetals}

\author{A. M. Marques}
\email{anselmomagalhaes@ua.pt}
\affiliation{Department of Physics $\&$ i3N, University of Aveiro, 3810-193 Aveiro, Portugal}

\author{R. G. Dias}
\affiliation{Department of Physics $\&$ i3N, University of Aveiro, 3810-193 Aveiro, Portugal}



\begin{abstract}
Recently, we have introduced in [A. M. Marques \textit{et al.}, Phys. Rev. B \textbf{103}, 235425 (2021)] the concept of $2^n$-root topology and applied it to one-dimensional (1D) systems.
These models require $n$ squaring operations to their Hamiltonians, intercalated with different constant energy downshifts at each level, in order to arrive at a decoupled block corresponding to a known topological insulator (TI) that acts as the source of the topological features of the starting $2^n$-root TI ($\sqrt[2^n]{\text{TI}}$).
In the process, $n$ non-topological residual models with degenerate spectra and in-gap impurity states appear, which dilute the topologically protected component of the starting edge states. 
Here, we generalize this method to several two-dimensional (2D) models, by finding the 4-root version of lattices hosting weak and higher-order boundary modes (both topological and non-topological) of a Chern insulator and of a topological semimetal.
We further show that a starting model with a non-Hermitian region in parameter space and a complex energy spectrum can nevertheless display a purely real spectrum for all its successive squared versions, allowing for an exact mapping between certain non-Hermitian models and their Hermitian lower root-degree counterparts.
A comment is made on the possible realization of these models in artificial lattices. 
\end{abstract}

\maketitle
\section{Introduction}
\label{sec:intro}

Square-root topology is one of the newest additions to the taxonomy of topological systems \cite{Hasan2010}.
It classifies a class of models whose conventional topological characterization, based on quantized topological invariants, fails when applied directly, but can be recovered when applied to the squared Hamiltonian.
An exact mapping between the eigenstates of the original and squared Hamiltonians allows one to track precisely how the topological features of the latter are inherited by the former \cite{Kremer2020}.
After its discovery \cite{Arkinstall2017}, a growing body of literature dedicated to  $\sqrt{\text{TIs}}$ has been developing \cite{Pelegri2019,Mizoguchi2020,Ezawa2020,Ke2020,Mizoguchi2020b,Lin2021}, with some reports providing experimental evidence for these models in different platforms already available \cite{Kremer2020,Song2020,Yan2020}.
Recently, a recipe for the identification of a new subset of $\sqrt{\text{TIs}}$ directly in reciprocal-space has been proposed \cite{Yoshida2021}, and a modified Altland-Zirbauer table \cite{Altland1997} for their topological classification, with all classes projected onto the A group of the conventional table \cite{Hasan2010}, was put forth.

\begin{figure}[ht]
	\begin{centering}
		\includegraphics[width=0.45 \textwidth,height=8.5cm]{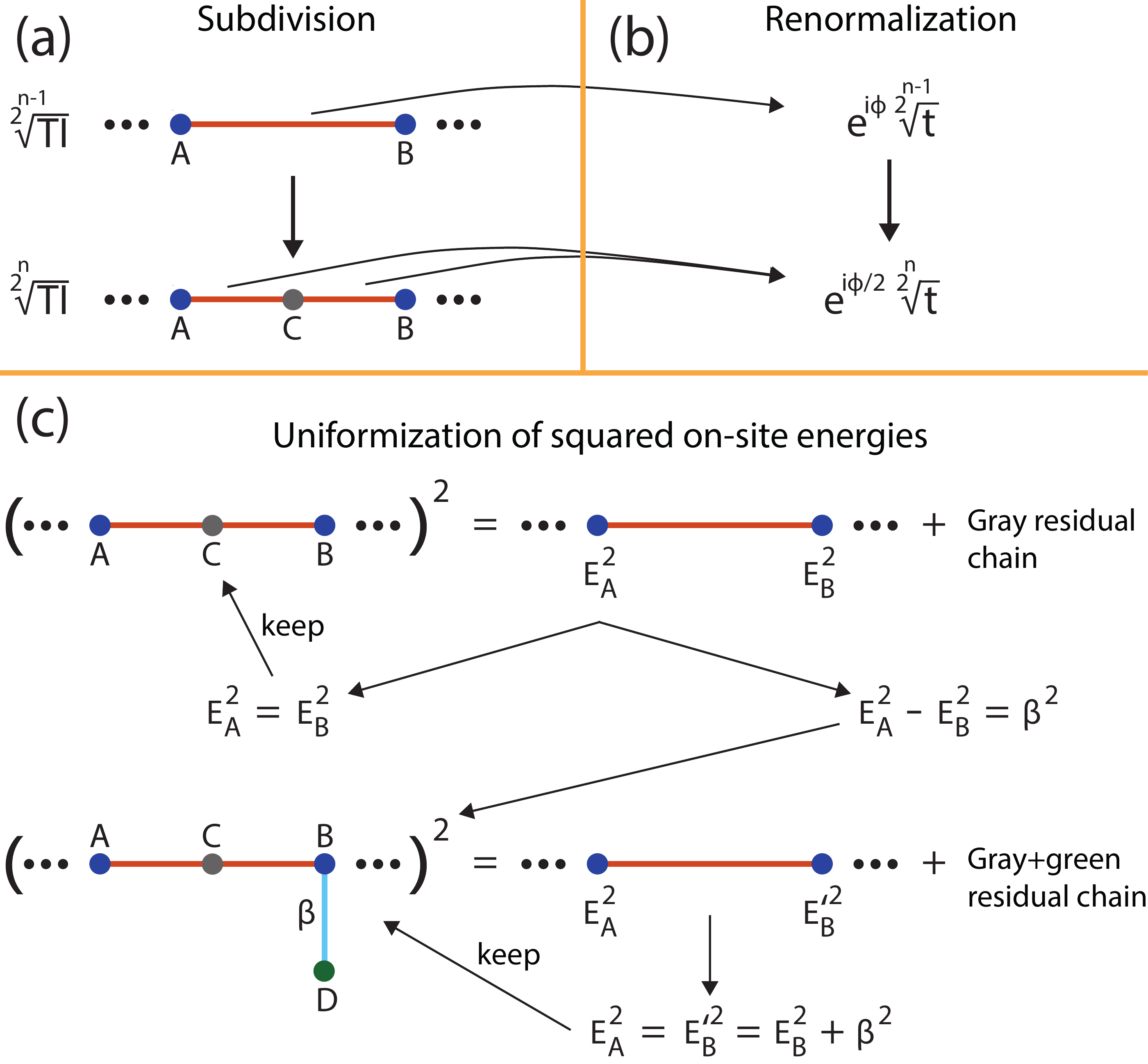} 
		\par\end{centering}
	\caption{Schematic depiction of the main ingredients for the construction of a higher root-degree version ($\sqrt[2^n]{\text{TI}}$) from its lower-root degree chain ($\sqrt[2^{n-1}]{\text{TI}}$). (a) Subdivision of the $\sqrt[2^{n-1}]{\text{TI}}$ by introducing a site in the middle of each hopping term. (b) Renormalization of the new doubled hopping terms, both in amplitude and in phase. (C) Uniformization of the squared on-site energy at the relevant blue sublattice the $\sqrt[2^{n}]{\text{TI}}$ which, if not achieved by default, requires the introduction of extra sites connected to the blue sites with lower squared on-site energies with properly tuned hopping strengths. Additional subtleties of this method can be found in [\onlinecite{Marques2021}] and here at the end of Sec.~\ref{sec:qrasym2dssh}.}
	\label{fig:mainsteps}
\end{figure}

The rules for the construction of square-root models have been introduced in [\onlinecite{Mizoguchi2020}] and systematized in [\onlinecite{Ezawa2020}], and heavily rely on the bipartition of the resulting square-root model, even when the original model is not bipartite itself.
The two main ingredients consist of subdividing the lattice by introducing additional sites in the middle of each link (hopping term) and renormalizing the magnitudes and phases of the doubled hopping terms, as schematically depicted in Figs.~\ref{fig:mainsteps}(a) and \ref{fig:mainsteps}(b).
In a previous work \cite{Marques2021}, we showed that the inclusion of an extra rule, invoked to ensure a uniform on-site potential at the sites in the topological block of the squared Hamiltonian [see Fig.~\ref{fig:mainsteps}(c)], enables an iterative application of the method such that the higher root-degree versions (the $\sqrt[2^n]{\text{TI}}$ with $n\geq 2$) can be derived from the $\sqrt{\text{TI}}$.
This method was applied there to one-dimensional (1D) models, including the Kitaev chain \cite{Kitaev2001}, a well-known 1D topological superconductor, and to a 1D model with self-similar properties after successive squaring operations in another work \cite{Dias2021}.

Here, we extend the results of [\onlinecite{Marques2021}] to several relevant two-dimensional (2D) models with different topological behavior.
More concretely, we systematically construct the quartic-root versions of a system with simultaneous weak and higher-order topological phases, a Chern insulator, a triangular lattice hosting conventional edge and corner modes, and a topological semimetal in a cylindrical geometry.
We further outline the general method for straightforwardly deriving the higher root-degree versions of these lattices.
The higher-order $\sqrt[2^n]{\text{TIs}}$ are seen to be characterized by the presence of finite-energy in-gap corner modes in their spectra with a fractional topological component inherited from the zero-energy corner states of the original TI.
The former have a weaker form of topological protection involving the chiral symmetry present at the sublattice of the original TI \cite{Marques2019}.
For the $\sqrt[2^n]{\text{TIs}}$ with $\mathbb{Z}_2$ topological indices, these are seen to display non-quantized values at the energy gaps hosting the boundary modes, and whose quantization is only recovered for the original TI obtained after $n$ squaring operations.

The rest of the paper is organized as follows.
In Sec.~\ref{sec:asym2dssh}, we introduce and characterize a 2D lattice hosting weak and higher-order topological states.
In Sec.~\ref{sec:srasym2dssh}, we carefully construct its square-root version, showing how the topological invariants lose their quantization and how it can be recovered by squaring the Hamiltonian in order to arrive at the original model.
In Sec.~\ref{sec:qrasym2dssh}, we go beyond square-root topology and construct the quartic-root version of the lattice model, and finish this section with the general recipe for the construction of the $2^n$-root version for any $n\geq 2$.
The following three sections are devoted to the application of the method outlined at the end of Section~\ref{sec:qrasym2dssh} to find the 4-root versions of different 2D systems, namely, a Chern insulator with chiral pairs of edge bands in Sec.~\ref{sec:sr4ci}, a triangular lattice model hosting edge and corner modes in Sec.~\ref{sec:sr4hoti}, and a semimetal  with topological flat bands of edge states in Sec.~\ref{sec:sr4ts}.
Finally, we conclude in Section~\ref{sec:conclusions}.

\section{Edge and corner states in the asymmetric 2D Su-Schrieffer-Heeger model}
\label{sec:asym2dssh}

\begin{figure}[ht]
	\begin{centering}
		\includegraphics[width=0.375 \textwidth,height=12cm]{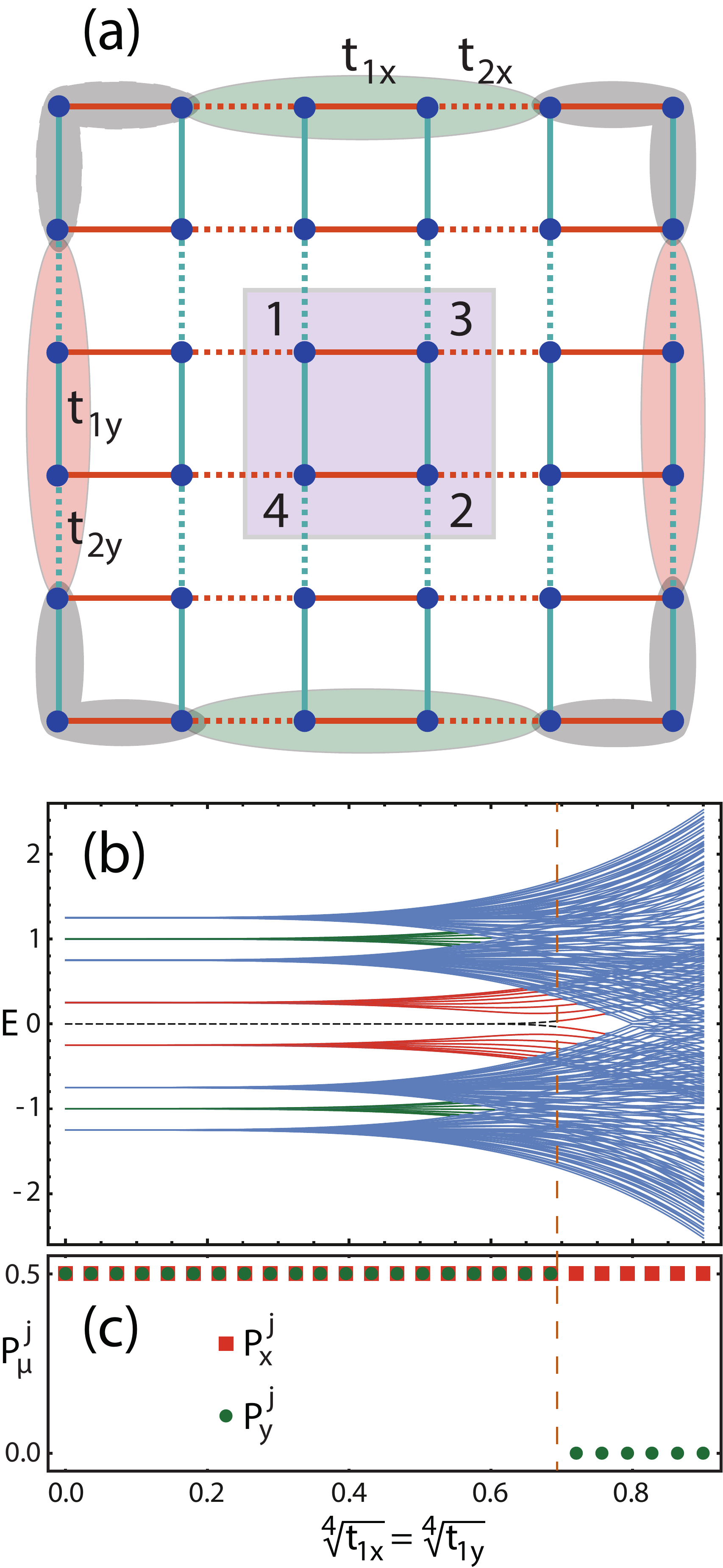} 
		\par\end{centering}
	\caption{(a) Asymmetric 2D SSH lattice. Red (green) shaded region indicates the localization of the horizontal (vertical) edge states, while the gray shaded region indicates the localization of the corner states. The purple shaded square in the middle delimits the unit cell. 
	(b) Energy spectrum for the model in (a) for an open square lattice of $8\times 8$ unit cells, with $t_{2x}=1$ (the energy unit) and $t_{2y}=0.25$, as a function of a linear variation of $\sqrt[4]{t_{1x}}=\sqrt[4]{t_{1y}}$. Solid blue curves correspond to bulk states, solid red (green) curves to horizontal (vertical) edge states and the four-fold degenerate dashed black zero-energy curve to the corner states.
    (c) Polarization profile, in units of $e$, along the $x$ (red squares) and $y$ (green dots) direction for the bulk band $j=1,2,3,4$ (it is the same for all bands).}
	\label{fig:hotibo}
\end{figure}

A simple model exhibiting a simultaneous weak and higher-order topological phase is provided by the asymmetric 2D Su-Schrieffer-Heeger (SSH) model, composed of coupled vertical and horizontal stacks of SSH chains \cite{Su1979}, with non-equivalent dimerized hoppings between orthogonal directions, as depicted in Fig.~\ref{fig:hotibo}(a).
The bulk Hamiltonian of this higher-order TI (HOTI), written in the ordered $\{\ket{j,\mathbf{k}}\}$ basis, with $j=1,2,3,4$, $\mathbf{k}=(k_x,k_y)$ and lattice constants set to $a_x=a_y\equiv 1$, is given by
\begin{eqnarray}
	H_{\text{HOTI}}(\mathbf{k})&=&
	\begin{pmatrix}
	0&h_{\text{HOTI}}^\dagger
	\\
	h_{\text{HOTI}}&0
	\end{pmatrix},
	\label{eq:hamilthoti}
	\\
	h_{\text{HOTI}}^\dagger&=&
	\begin{pmatrix}
	t_{1x}+t_{2x}e^{-ik_x}&t_{1y}+t_{2y}e^{ik_y}
	\\
	t_{1y}+t_{2y}e^{-ik_y}&t_{1x}+t_{2x}e^{ik_x}
	\end{pmatrix},
\end{eqnarray}
where we assume $t_{i\mu}\geq 0$, with $i=1,2$ and $\mu=x,y$.
Diagonalization of this Hamiltonian yields four energy bands,
\begin{eqnarray}
E_1(\mathbf{k})&=&-E_4(\mathbf{k})=-D_x(k_x)-D_y(k_y),
\label{eq:e1e4}
\\
E_2(\mathbf{k})&=&-E_3(\mathbf{k})=-D_x(k_x)+D_y(k_y),
\label{eq:e2e3}
\end{eqnarray}
with $D_\mu(k_\mu)=\sqrt{t_{1\mu}^2+t_{2\mu}^2+2t_{1\mu}t_{2\mu}\cos k_\mu}$.

In the Benalcazar-Bernevig-Hughes (BBH) model \cite{Benalcazar2017,Benalcazar2017b}, consisting of a 2D SSH model \cite{Liu2017} with a $\pi$-flux per plaquette, the flux is responsible for opening a zero-energy gap where corner states can be found in the topological phase under open boundary conditions (OBC).
Without the inserted flux, the corner modes form bound states in the continuum buried within the bulk bands at zero-energy \cite{Pino2020}, which were recently shown to keep their topological protection, given that they do not hybridize with the surrounding bulk modes if chiral symmetry is preserved \cite{Cerjan2020,Wang2021b}.
Instead of inserting a finite flux per plaquette \cite{Zuo2021}, a gap between the two middle bands can be opened at zero flux either by including next-nearest-neighbor (NNN) hoppings \cite{Ivanova2020,Olekhno2021} or setting asymmetric hopping dimerizations between the $x$ and $y$ directions, as we are considering here.

\begin{figure}[ht]
	\begin{centering}
		\includegraphics[width=0.475 \textwidth,height=5cm]{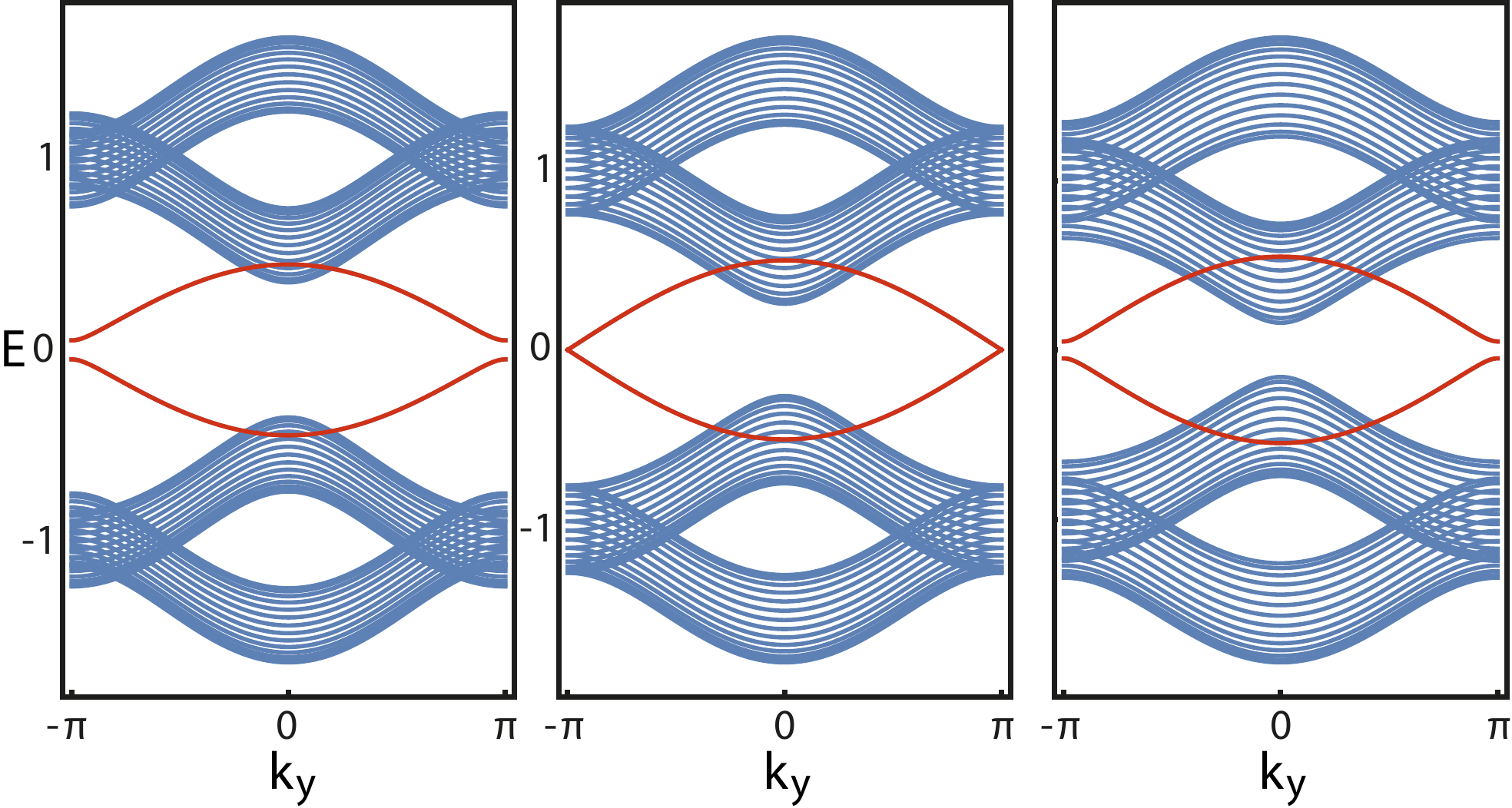} 
		\par\end{centering}
	\caption{Triptych of energy spectra for the asymmetric 2D SSH model open along $x$ and periodic along $y$ with 15 supercells, as a function of $k_y$. 
		Parameters: $t_{1x}=1$, $t_{2y}=0.25$ and $t_{1x}=t_{1y}=0.2$ for the left panel, $t_{1x}=t_{1y}=0.25$ for the middle panel and $t_{1x}=t_{1y}=0.3$ for the right panel.
	From one side to the other there is a gap closing and reopening between the two red bands of horizontal edge states.}
	\label{fig:ribbon}
\end{figure}

The energy spectrum for the model in Fig.~\ref{fig:hotibo}(a) for an open square lattice of $8\times 8$ unit cells, with $t_{2x}=1$ (the energy unit) and $t_{2y}=0.25$, as a function of a linear variation of $\sqrt[4]{t_{1x}}=\sqrt[4]{t_{1y}}$ (for reasons that will become clear below), is shown in Fig.~\ref{fig:hotibo}(b).
All gaps are open between the bulk bands (blue curves) for $\sqrt[4]{t_{1x}}\lesssim0.6$, with 1D vertical edge states \footnote{We label the edge states vertical or horizontal according to the direction over which they decay to the bulk.} (green curves) lying within the top and bottom gaps, and 1D horizontal edge states (symmetric red curves) in the middle gap which, simultaneously, hosts four degenerate zero-energy corner states (dashed black curves). 
Notice that the corner states disappear when the gap between the horizontal edge states closes, at $t_{1x}=t_{2y}=0.25$ (or $\sqrt[4]{t_{1x}}\approx 0.7$), \textit{before} the bulk gap closing at $\sqrt[4]{t_{1x}}\approx 0.8$.
This closing and reopening of the middle edge bands is highlighted in the triptych of energy spectra in Fig.~\ref{fig:ribbon}, where a ribbon geometry open in $x$ and periodic in $y$ is considered.
Such higher-order states, whose topological phase is protected by a boundary energy gap have been recently termed extrinsic \cite{Geier2018} or  boundary-obstructed TIs \cite{Khalaf2021,Ezawa2020c}.
The corner modes cannot be gapped out by chiral-symmetry preserving perturbations, both at the bulk and boundaries, as long as the energy gap between the bands of horizontal edge states remains open.

The polarization vector per unit length for each band $j$, $\mathbf{P}^j=(P_x^j,P_y^j)$, serves as the weak topological index of the system characterizing the 1D edge states, and can be computed from
\begin{eqnarray}
P_\mu^j&=&\frac{e}{2\pi}\gamma_\mu^j,
\label{eq:polarizationj}
\\
\gamma_\mu^j&=&\frac{i}{2\pi}\int_{BZ}d\mathbf{k}\bra{u_j(\mathbf{k})}\frac{d}{dk_{\mu}}\ket{u_j(\mathbf{k})},
\label{eq:2dzak}
\end{eqnarray}
where $e$ is the electron charge, $\gamma_\mu^j$ is the 2D Zak's phase \cite{Delplace2011} for the $\mu=x,y$ direction defined modulo $2\pi$, the integral goes over the 2D Brillouin zone and $\ket{u_j(\mathbf{k})}$ is the eigenvector of band $j$.
The polarization along each direction is quantized to $P_\mu^j=0,\frac{e}{2}$ by the presence of a ($\mathbf{k}$-independent) reflection symmetry along $\mu$, whose explicit expressions are given by
\begin{eqnarray}
M_xH_{\text{HOTI}}(k_x,k_y)M_x^{-1}&=&H_{\text{HOTI}}(-k_x,k_y),
\\ M_x&=&\sigma_x\otimes\sigma_0,
\label{eq:reflectionx}
\\
M_yH_{\text{HOTI}}(k_x,k_y)M_y^{-1}&=&H_{\text{HOTI}}(k_x,-k_y),
\\ 
 M_y&=&\sigma_x\otimes\sigma_x,
\label{eq:reflectiony}
\end{eqnarray}
with $\sigma_x$ being the $x$ Pauli matrix and $\sigma_0$ the $2\times 2$ identity matrix.
A successive application of the reflection symmetries yields a two-fold rotation symmetry written as
\begin{eqnarray}
C_2H_{\text{HOTI}}(\mathbf{k})C_2^{-1}&=&H_{\text{HOTI}}(-\mathbf{k}),
\\
C_2=M_xM_y&=&\sigma_0\otimes\sigma_x.
\end{eqnarray}
All four bands exhibit the polarization spectrum of Fig.~\ref{fig:hotibo}(c).
Naively, one might be tempted to define the polarization along $\mu$ at each gap by the usual expression
\begin{equation}
	P_\mu=\sum\limits_{j\in ``occ"}P_\mu^j \mod e,
	\label{eq:polarization}
\end{equation} 
where $``occ"$ is the set of occupied bands lying below the gap (assuming a spinless fermionic picture).
However, noting from (\ref{eq:e1e4})-(\ref{eq:e2e3}) that the asymmetric 2D SSH model is factorizable in the $x$ and $y$ directions, such that $\ket{u_j(\mathbf{k})}=\ket{u_{j,x}(k_x)}\otimes\ket{u_{j,y}(k_y)}$, then (\ref{eq:2dzak}) can be reduced to a 1D Zak's phase \cite{Zak1989},
\begin{equation}
\gamma_\mu^j=i\int_{BZ_\mu}dk_\mu\bra{u_{j,\mu}(k_\mu)}\frac{d}{dk_{\mu}}\ket{u_{j,\mu}(k_\mu)},
\end{equation}
where BZ$_\mu$ is the 1D Brillouin zone along $\mu$. 
In other words, $\gamma_\mu^j$ is $\bar{\mu}$ independent, where $\bar{\mu}$ is the direction orthogonal to $\mu$.
It follows that the pair of bands with energies $\pm D_\mu(k_\mu)-D_{\bar{\mu}}(k_{\bar{\mu}})$ can be decoupled from the pair with energies $\pm D_\mu(k_\mu)+D_{\bar{\mu}}(k_{\bar{\mu}})$ [see (\ref{eq:e1e4}) and (\ref{eq:e2e3})], such that each pair defines an energy gap whose topological characterization is \textit{independent} to that of the other pair.
Specifically, the computation of the polarization $P_x$ ($P_y$) considers each band pair $\{E_1,E_3\}$ and $\{E_2,E_4\}$ ($\{E_1,E_2\}$ and $\{E_3,E_4\}$) independently, where only the lowest band of the pair is considered occupied when the gap is opened between the two, in which case (\ref{eq:polarization}) can be recast as
\begin{equation}
	P_\mu^{\{i,j\}}=P_\mu^i,\ \ i<j,
\end{equation}
where $\{i,j\}$ is the pair index.
In agreement with Fig.~\ref{fig:hotibo}(c), one readily finds
\begin{eqnarray}
P_x^{\{1,3\}}&=&P_x^{\{2,4\}}=
\begin{cases}
\frac{e}{2},\ \  \sqrt[4]{t_{1x}}<1,
\\
0, \ \ \sqrt[4]{t_{1x}}>1,
\end{cases}
\\
P_y^{\{1,2\}}&=&P_y^{\{3,4\}}=
\begin{cases}
\frac{e}{2},\ \  \sqrt[4]{t_{1x}}\lesssim 0.7,
\\
0, \ \ \sqrt[4]{t_{1x}}\gtrsim 0.7,
\end{cases}
\end{eqnarray}
where, in the topological phase and under OBC, the non-trivial $P_x^{\{1,3\}}$ ($P_x^{\{2,4\}}$) has a correspondence with the lower (higher) red band of horizontal edge states in Fig.~\ref{fig:hotibo}(b), whereas the non-trivial $P_y^{\{1,2\}}$ ($P_y^{\{3,4\}}$) has a correspondence with the lower (higher) green band of vertical edge states.

The four zero-energy corner states, given by the dashed black curves in Fig.~\ref{fig:hotibo}(b), are present until $t_{1x}=0.25$ ($\sqrt[4]{t_{1x}}\approx 0.7$), which marks the point where the gap between the horizontal red edge bands closes and, at the same time, the point where the bulk gap between both the two lower and higher bands closes, that is, the point where the vertical green edge states vanish into the bulk. 
Clearly, the corner states are a consequence of the confluence of $x$ and $y$ edge polarizations.
The quadrupole moment per unit area, $q_{xy}$, can be identified as the higher-order $\mathbb{Z}_2$ topological index characterizing the presence of absence or corner states under OBC.
However, the formula of $q_{xy}$ used for the BBH model \cite{Benalcazar2017,Benalcazar2017b} cannot be used here, as it relies on the non-commutativity of the reflection symmetries, whereas we have $[M_x,M_y]=0$ from (\ref{eq:reflectionx}) and (\ref{eq:reflectiony}).
An alternative formula for an \textit{ad hoc} computation of $q_{xy}$ for the middle energy gap of the asymmetric 2D SSH model was provided in Ref.~[\onlinecite{Pelegri2019c}], and reads as
\begin{eqnarray}
	q_{xy}&=&\frac{1}{2e}\tilde{P}_x\tilde{P}_y,
	\label{eq:4moment}
	\\
	\tilde{P}_\mu&=&\sum\limits_{j=1,2}P_\mu^j,
\end{eqnarray}
from where we find
\begin{equation}
	q_{xy}=
\begin{cases}
\frac{e}{2},\ \  \sqrt[4]{t_{1x}}\lesssim 0.7,
\\
0, \ \ \sqrt[4]{t_{1x}}\gtrsim 0.7,
\end{cases}
\end{equation}
and the corresponding presence (absence) of corner states under OBC for the parameter region with non-trivial (trivial) $q_{xy}$ can be checked against Fig.~\ref{fig:hotibo}(b).

\section{Square root of the asymmetric 2D SSH model}
\label{sec:srasym2dssh}

\begin{figure}[ht]
	\begin{centering}
		\includegraphics[width=0.375 \textwidth,height=12cm]{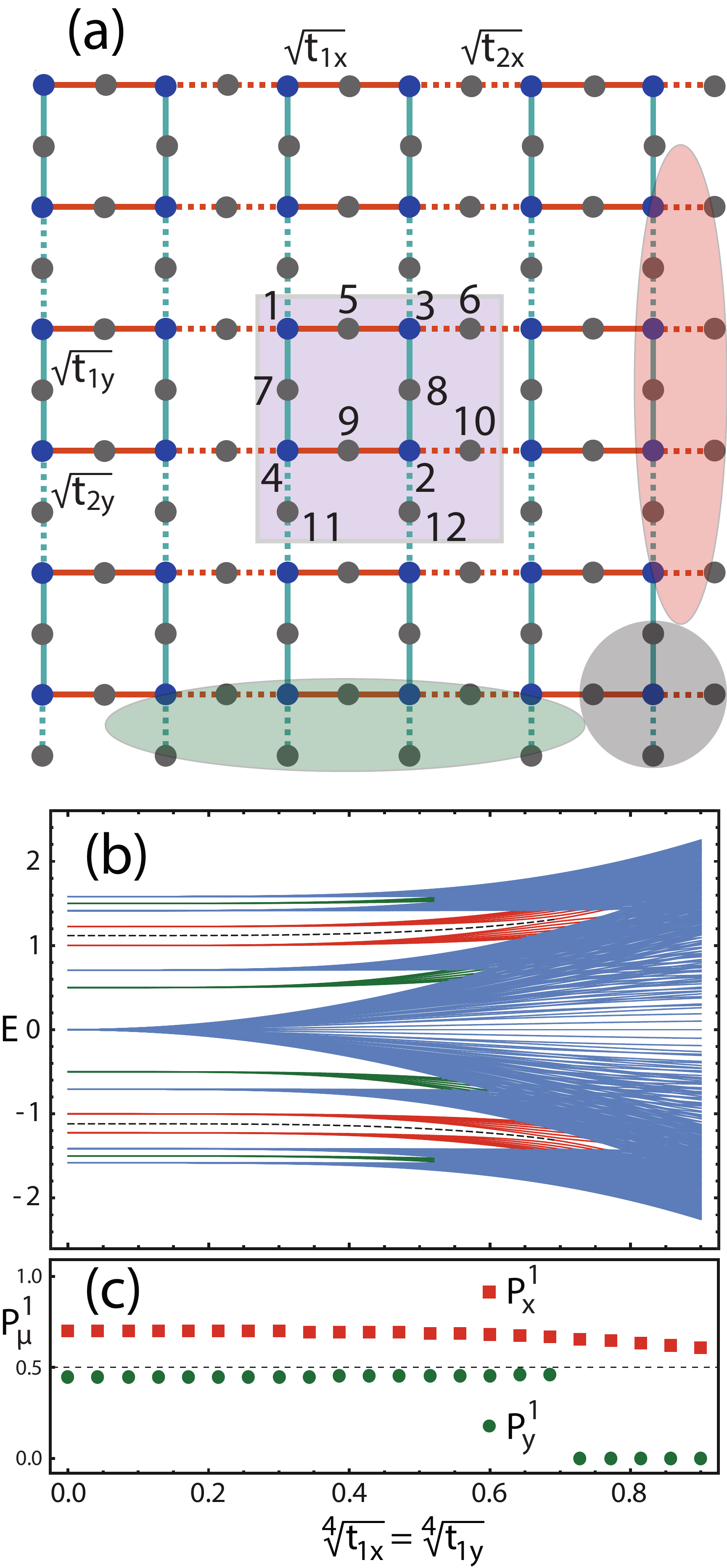} 
		\par\end{centering}
	\caption{(a) Square-root version of the asymmetric 2D SSH lattice. The meaning of the shaded regions is explained in Fig.~\ref{fig:hotibo}(a).
		(b) Energy spectrum for the model in (a) for an open square lattice of $8\times 8$ unit cells for the same parameters and color scheme for the curves as in Fig.~\ref{fig:hotibo}(b).
		(c) Polarization profile, in units of $e$, along the $\mu=x,y$ direction for the lowest bulk band.
	    The dashed line at $P_\mu^1=0.5$ is a guide to the eye.}
	\label{fig:hotibo2}
\end{figure}

The square-root version of the asymmetric 2D SSH model can be constructed by direct application of the recipe in Ref.~[\onlinecite{Ezawa2020}]: firstly, one constructs the split graph \cite{Ma2020} of the model in Fig.~\ref{fig:hotibo}(a) by including a site at the middle of each link representing a hopping term, such that each link gets divided in two; secondly, the correspondence between the original and square-root versions of a model is kept by renormalizing each pair of subdivided hopping terms, in relation to the common original one, as $t_{i\mu}\to \sqrt{t_{i\mu}}$.
These two steps lead to the model in Fig.~\ref{fig:hotibo2}(a), where the new sites created by subdivision of the original model form the gray sublattice and the blue sites form the sublattice corresponding to the sites in the original asymmetric 2D SSH lattice.
After this procedure, an ambiguity manifests itself regarding the definition of the unit cell, with the choice in Fig.~\ref{fig:hotibo2}(a) being one among several others (e.g., sites 6 and 10 could be placed at the left in the unit cell and/or sites 11 and 12 could be placed at the top in the unit cell).
A way to neutralize this ambiguity in real space, which will be adopted in the next sections, is by including extra sites along the edges such that all choices of unit cell become equivalent.
In the case of Fig.~\ref{fig:hotibo2}(a), this would translate in adding extra gray sites 6 and 10 along the left edge and sites 11 and 12 along the top edge.

The bulk Hamiltonian of the model in Fig.~\ref{fig:hotibo2}(a), written in the ordered $\{\ket{j,\mathbf{k}}\}$ basis, with $j=1,2,\dots,12$, is given by
\begin{eqnarray}
H_{\sqrt{\text{HOTI}}}(\mathbf{k})&=&
\begin{pmatrix}
0&h_{\sqrt{\text{HOTI}}}^\dagger
\\
h_{\sqrt{\text{HOTI}}}&0
\end{pmatrix},
\label{eq:hamilthotibo2}
\\
h_{\sqrt{\text{HOTI}}}&=&
\begin{pmatrix}
\sqrt{t_{1x}}&0&\sqrt{t_{1x}}&0
\\
\sqrt{t_{2x}}e^{ik_x}&0&\sqrt{t_{2x}}&0
\\
\sqrt{t_{1y}}&0&0&\sqrt{t_{1y}}
\\
0&\sqrt{t_{1y}}&\sqrt{t_{1y}}&0
\\
0&\sqrt{t_{1x}}&0&\sqrt{t_{1x}}
\\
0&\sqrt{t_{2x}}&0&\sqrt{t_{2x}}e^{ik_x}
\\
\sqrt{t_{2y}}e^{-k_y}&0&0&\sqrt{t_{2y}}
\\
0&\sqrt{t_{2y}}&\sqrt{t_{2y}}e^{-ik_y}&0
\end{pmatrix}.
\end{eqnarray}
In contrast with $H_{\text{HOTI}}$ in (\ref{eq:hamilthoti}), the spatial symmetries of $H_{\sqrt{\text{HOTI}}}$ are $\mathbf{k}$ dependent, as a consequence of non-centered reflection-axes within the unit cell \cite{Marques2019},
\begin{eqnarray}
	M_x&=&\begin{pmatrix}
	M_x^{\text{BS}}&0
	\\
	0&M_x^{\text{GS}}
	\end{pmatrix},
	\label{eq:mx2}
	\\
	M_x^{\text{BS}}&=&\sigma_x\otimes\sigma_0,
	\\
	M_x^{\text{GS}}&=&\text{diag}\big(1,e^{-ik_x},\sigma_x,1,e^{-ik_x},\sigma_x\big),
\end{eqnarray}
and
\begin{eqnarray}
M_y&=&\begin{pmatrix}
M_y^{\text{BS}}&0
\\
0&M_y^{\text{GS}}
\end{pmatrix},
\label{eq:my2}
\\
M_y^{\text{BS}}&=&\sigma_x\otimes\sigma_x,
\\
M_y^{\text{GS}}&=&
\begin{pmatrix}
0&0&0&0&1&0&0&0
\\
0&0&0&0&0&1&0&0
\\
0&0&1&0&0&0&0&0
\\
0&0&0&1&0&0&0&0
\\
1&0&0&0&0&0&0&0
\\
0&1&0&0&0&0&0&0
\\
0&0&0&0&0&0&e^{ik_y}&0
\\
0&0&0&0&0&0&0&e^{ik_y}
\end{pmatrix},
\end{eqnarray}
where BS (GS) stands for blue (gray) sublattice.
According to the Lieb's theorem \cite{Lieb1989}, the sublattice imbalance within the unit cell ($\#_{GS}-\#_{BS}=4$) equals the number of zero-energy flat bands under PBC.
This is the mechanism behind the formation of highly degenerate zero-energy flat bands that appear for a class of Lieb-like lattices studied recently \cite{Ni2020}.
\begin{figure*}[ht]
	\begin{centering}
		\includegraphics[width=0.9 \textwidth,height=5.5cm]{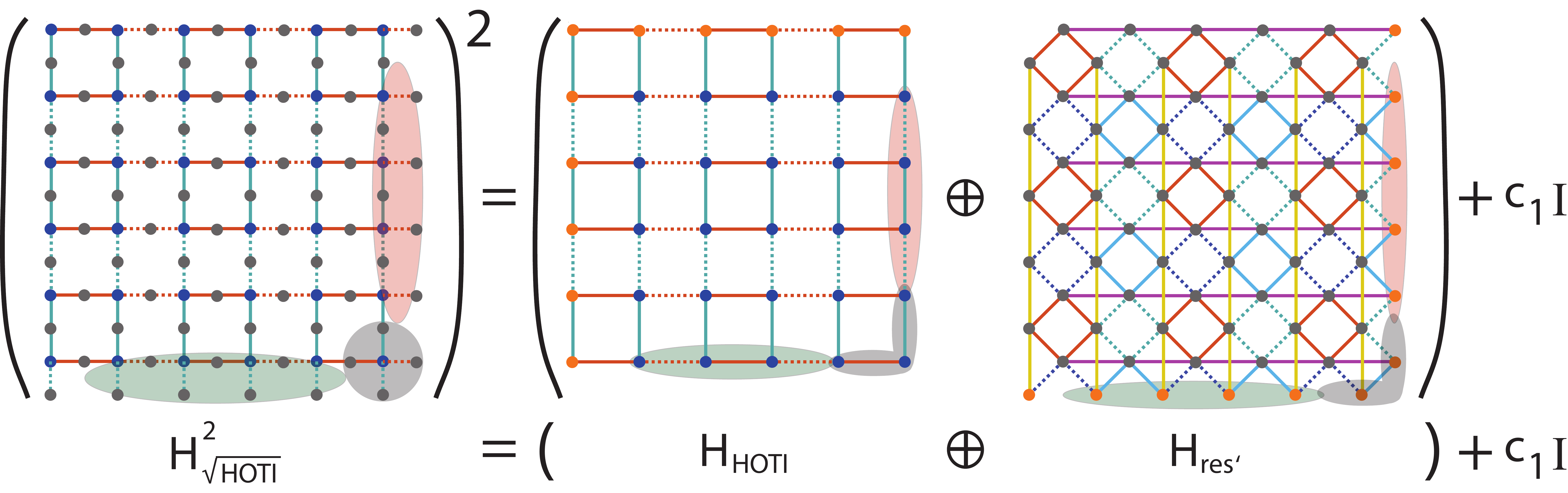} 
		\par\end{centering}
	\caption{Schematic illustration of the result of squaring the $\sqrt{\text{HOTI}}$ lattice (and the respective Hamiltonian below) with an integer number of unit cells in $x$ and $y$. Up to a constant overall energy shift $c_1$,the squared model is composed of two decoupled lattices, one yielding the HOTI in the blue sublattice and the other yielding the residual model in the gray sublattice. The meaning of the shaded areas is given in Fig.~\ref{fig:hotibo}(a). Orange impurity sites with an on-site energy offset in relation to the other sites of the same lattice appear at different boundaries with opposite effects: they gap out the boundary modes in the HOTI lattice but drive the boundary modes in the residual lattice to the same gaps of the corresponding ones in the HOTI.}
	\label{fig:squaredhotibo2}
\end{figure*}
The combination of the reflection symmetries leads to a two-fold rotation symmetry written as
\begin{eqnarray}
C_2&=&M_xM_y=\begin{pmatrix}
C_2^{BS}&0
\\
0&C_2^{GS}
\end{pmatrix},
\label{eq:c22}
\\
C_2^{BS}&=&\sigma_0\otimes\sigma_x,
\\
C_2^{GS}&=&
\begin{pmatrix}
0&0&0&0&1&0&0&0
\\
0&0&0&0&0&e^{-1k_x}&0&0
\\
0&0&0&1&0&0&0&0
\\
0&0&1&0&0&0&0&0
\\
1&0&0&0&0&0&0&0
\\
0&e^{-1k_x}&0&0&0&0&0&0
\\
0&0&0&0&0&0&0&e^{ik_y}
\\
0&0&0&0&0&0&e^{ik_y}&
\end{pmatrix}.
\end{eqnarray}
The $k_\mu$-dependence in $M_\mu$ implies that $\gamma_\mu^j$ in (\ref{eq:2dzak}) is not $\pi$-quantized in general \cite{Marques2019}.
In turn, neither the polarization $P_\mu^j$ in (\ref{eq:polarizationj}) nor, consequently, the quadrupole moment $q_{xy}$ in (\ref{eq:4moment}), regardless of the energy gap considered, will necessarily be quantized, as exemplified by the  polarization profile of the lowest energy band shown in Fig.~\ref{fig:hotibo2}(c).
Quantized topological indices will only be recovered at the level of the diagonal block of the squared model corresponding to the asymmetric 2D SSH model analyzed in the previous section.
This truly establishes the model in Fig.~\ref{fig:hotibo2}(a) as a square-root HOTI, since the usual tools for the topological characterization of its boundary modes fail when directly applied, and must be derived from the ``well behaved'' topological features of the original HOTI one arrives at by squaring the model.

The $\mathbf{k}$-dependence of the bulk spatial symmetries \cite{Madail2019} [see (\ref{eq:mx2}), (\ref{eq:my2}) and (\ref{eq:c22})], regardless of the choice of unit cell, means that these same symmetries are broken in real-space when an \textit{integer} number of unit cell is considered, as in Fig.~\ref{fig:hotibo2}(a).
In order to restore the same symmetries in real-space, extra sites from incomplete unit cells would have to be included along the two edges where they are missing [e.g, including the dangling gray sites present at the right and bottom edges also along the left and top edges in Fig.~\ref{fig:hotibo2}(a), as explained above].
The energy spectrum for the model in Fig.~\ref{fig:hotibo2}(a) for an open square lattice of $8\times 8$ unit cells, with $t_{2x}=1$ (the energy unit) and $t_{2y}=0.25$, keeping it, as before, as a function of a linear variation of $\sqrt[4]{t_{1x}}=\sqrt[4]{t_{1y}}$, is shown in Fig.~\ref{fig:hotibo2}(b).
The two corner bands form a finite energy chiral pair, with their states appearing at the bottom right corner as illustrated in Fig.~\ref{fig:hotibo2}(a).
Four green (red) bands of vertical (horizontal) edge states are present until the bulk gap closing points occurring at $\sqrt[4]{t_{1x}}\approx 0.7$ ($\sqrt[4]{t_{1x}}=1$) are reached, and are located around a single edge, more concretely the bottom (right) edge with the geometry of Fig.~\ref{fig:hotibo2}(a).
Again, the corner states evolve into horizontal edge states across the gap closing point between pairs of horizontal edge bands occurring at $\sqrt[4]{t_{1x}}=\sqrt[4]{0.25}\approx 0.7$ (that is, $\sqrt{t_{1x}}=0.5$, if we use the square-root energy units of the $\sqrt{\text{HOTI}}$ model of this section).

Squaring the $\sqrt{\text{HOTI}}$ Hamiltonian in (\ref{eq:hamilthotibo2}) leads to
\begin{eqnarray}
H^2_{\sqrt{\text{HOTI}}}(\mathbf{k})&=&
\begin{pmatrix}
H_{\text{par}}&0
\\
0&H_{\text{res}}
\end{pmatrix},
\label{eq:hamilthotibo22}
\\
H_{\text{par}}&=&h_{\sqrt{\text{HOTI}}}^\dagger h_{\sqrt{\text{HOTI}}}=c_1 I_4+H_{\text{HOTI}},
\label{eq:hamiltpar}
\\
H_{\text{res}}&=&h_{\sqrt{\text{HOTI}}} h_{\sqrt{\text{HOTI}}}^\dagger=c_1 I_8+H_{\text{res}^\prime},
\label{eq:hamiltres}
\end{eqnarray}
where $H_\text{par}$ ($H_\text{res}$) is the parent (residual) Hamiltonian, $H_\text{HOTI}$ is given in (\ref{eq:hamilthoti}), $c_1=t_{1x}+t_{2x}+t_{1y}+t_{2y}$ is a constant energy shift, $I_n$ is the $n\times n$ identity matrix.
The result of squaring the open lattice in Fig.~\ref{fig:hotibo2}(a) is schematically depicted in Fig.~\ref{fig:squaredhotibo2}, where it can be seen that edge and corner states are located around the unperturbed (perturbed) edges and corners, respectively, for the lattice corresponding to $H_\text{HOTI}$ ($H_{\text{res}^\prime}$).
The different offsets in the self-energies of the perturbed sites, all colored in orange for simplicity, are a consequence of their lower coordination number in the $\sqrt{\text{HOTI}}$ lattice, in relation to the other sites in the same sublattice.
Due to this, when $H_{\sqrt{\text{HOTI}}}$ is squared in real-space, and for an integer number of unit cells, these sites acquire fewer terms in their self-energy, akin to the edge corrections appearing in second-order perturbation theory to the states with weight at the edge sites \cite{Marques2017}.
As such, the boundary modes coming from $H_\text{HOTI}$ are topologically originated, as demonstrated in the previous section, while the ones coming from $H_{\text{res}^\prime}$ are impurity (also called Tamm \cite{Tamm1932} or Shockley \cite{Shockley1939}) states forming a \textit{degenerate} boundary spectrum with the former.
The energy spectrum coming from the diagonalization of $H^2_{\sqrt{\text{HOTI}}}$ with $8\times 8$ unit cells and a global down energy shift of $c_1(t_{1x})=1.25+2t_{1x}$ (since $t_{2x}=1$, $t_{2y}=0.25$, and $t_{1x}=t_{1y}$), coincides with the one in Fig.~\ref{fig:hotibo}(b), with all boundary curves doubly degenerate, both edge and corner ones.
From (\ref{eq:hamiltpar}) and (\ref{eq:hamiltres}) one has \cite{Ezawa2020}
\begin{eqnarray}
	H_\text{HOTI}\ket{\psi^\text{HOTI}_j}&=&(\epsilon_j-c_1)\ket{\psi^\text{HOTI}_j},
    \\
	H_\text{par}\ket{\psi^\text{HOTI}_j}&=&\epsilon_j\ket{\psi^\text{HOTI}_j},
	\\
	H_{\text{res}^\prime}\ket{\psi^{\text{res}^\prime}_l}&=&(\epsilon_l^{res}-c_1)\ket{\psi^{\text{res}^\prime}_l},
	\\
	H_\text{res}\ket{\psi^{\text{res}^\prime}_l}&=&\epsilon_l^{res}\ket{\psi^{\text{res}^\prime}_l},
\end{eqnarray}
where $\ket{\psi^\text{HOTI}_j}$ ($\ket{\psi^{\text{res}^\prime}_l}$) is the $j^{\text{th}}$ ($l^{\text{th}}$) eigenstate of the parent (residual) Hamiltonian, $\{\epsilon_j\}=\{\epsilon_1,\dots,\epsilon_{4N}\}$ with $\epsilon_j\geq 0$, $N$ is the total number of unit cells, $\{\epsilon_l^\text{res}\}=\{\{\epsilon_j\},\{0\}_{4N}\}$ and $\{0\}_{4N}=\{0,0,\dots,0\}$ is a vector of size $4N$ containing the excess zero-energy levels in $H_\text{res}$ coming from the sublattice imbalance in $H_{\sqrt{\text{HOTI}}}$.
Focusing on the finite energy states within the relevant set $\{\epsilon_j\}$, the eigenstates of the original $\sqrt{\text{HOTI}}$ lattice can be written in terms of the eigenstates of the HOTI and shifted residual lattices as
\begin{equation}
	H_{\sqrt{\text{HOTI}}}\ket{\psi_{j,\pm}}=\pm\sqrt{\epsilon_j}\ket{\psi_{j,\pm}})=\pm\sqrt{\epsilon_j}
	\begin{pmatrix}
	\ket{\psi^\text{HOTI}_j}
	\\
	\pm\ket{\psi^{\text{res}^\prime}_j}
	\end{pmatrix},
\end{equation}
where, due to the presence of chiral symmetry, defined in real-space as $CH_{\sqrt{\text{HOTI}}}C^{-1}=-H_{\sqrt{\text{HOTI}}}$, with $C=\text{diag}\big(I_{4N},-I_{8N}\big)$, all finite-energy states come in chiral pairs $\ket{\psi_{j,\pm}}$ (note, however, that the presence of impurity sites in the parent and residual lattices, as shown in Fig.~\ref{fig:squaredhotibo2}, breaks the chiral symmetry in $H_{\sqrt{\text{HOTI}}}^2-c_1I_{12}$).
In particular, both the weak edge and the higher-order corner states of the $\sqrt{\text{HOTI}}$ lattice in Fig.~\ref{fig:hotibo2}(b) can be said to be ``half''-topological and ``half''-impurity, with their topological features directly inherited from the $\ket{\psi^\text{HOTI}_j}$ component.
Thus, the model in Fig.~\ref{fig:hotibo2}(a) constitutes an example of a square-root weak and higher-order topological insulator.

\section{Quartic-root of the asymmetric 2D SSH model}
\label{sec:qrasym2dssh}
\begin{figure}[ht]
	\begin{centering}
		\includegraphics[width=0.475 \textwidth,height=6cm]{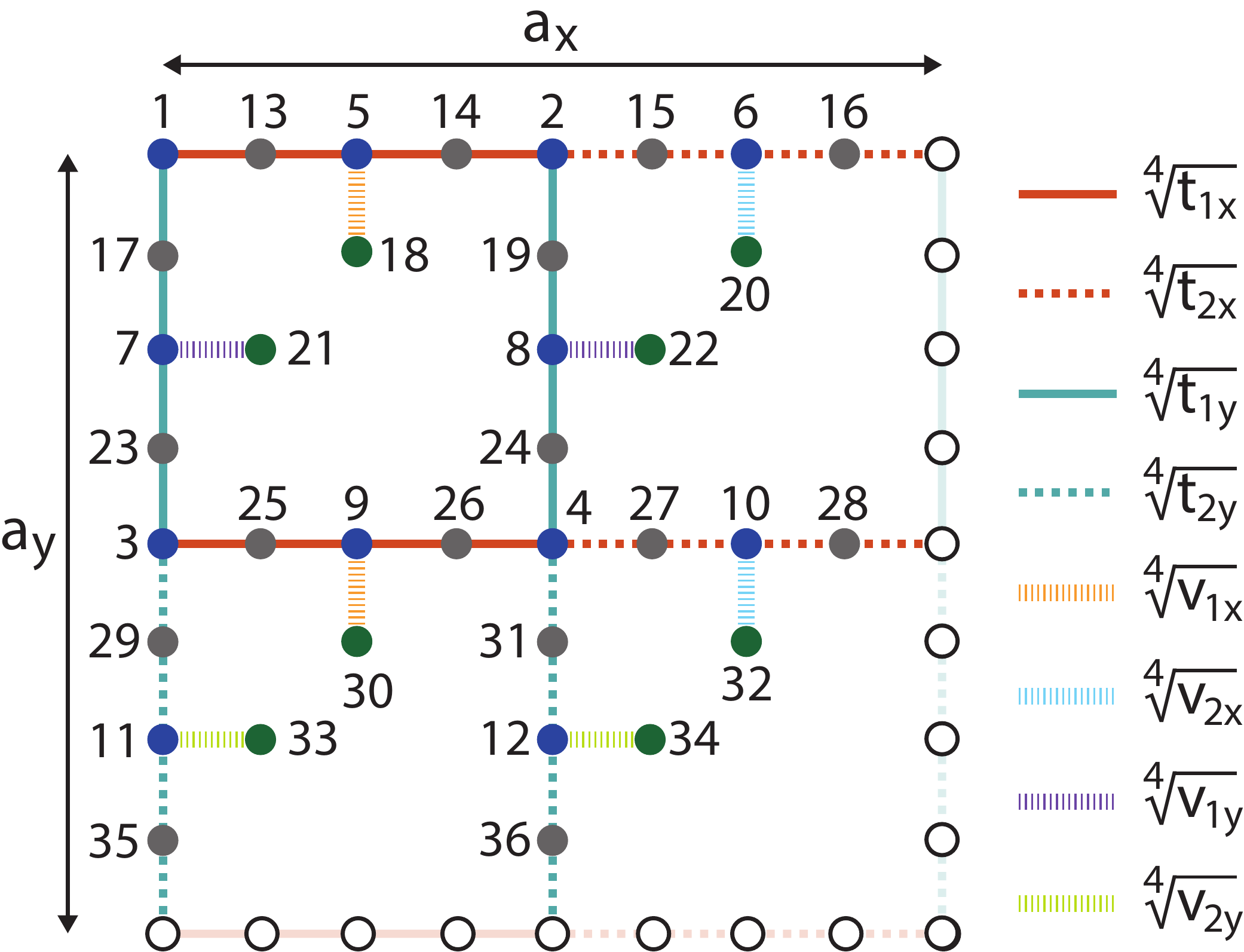} 
		\par\end{centering}
	\caption{Unit cell of the $\sqrt[4]{\text{HOTI}}$. Blue sites form one sublattice and the green and gray sites form another. Open sites at the edges belong to adjacent unit cells.}
	\label{fig:ucellboti4}
\end{figure}

Here, we want to go a step further than in the previous section and construct the quartic-root version of the asymmetric 2D SSH model, which we label $\sqrt[4]{\text{HOTI}}$.
Our starting point is the $\sqrt{\text{HOTI}}$ in Fig.~\ref{fig:hotibo2}, with all its sites colored in blue, since we will require that the 12 sites of its unit cell become a sublattice of the $\sqrt[4]{\text{HOTI}}$.
Then, as before, we subdivide the $\sqrt{\text{HOTI}}$ by including a gray site at the middle of each link and renormalizing the two resulting hoppings to the adjacent blue sites as $\sqrt{t_{i\mu}}\to \sqrt[4]{t_{i\mu}}$, leading to the unit cell depicted in Fig.~\ref{fig:ucellboti4} \textit{without} the green sites at this stage.
To see why this is not enough, we recall that the $\sqrt[4]{\text{HOTI}}$ model should be such that, upon squaring its bulk Hamiltonian, one gets the $\sqrt{\text{HOTI}}$ as one of its diagonal blocks, apart from a constant energy term, replicating for a higher root-degree the same relation established above between the $\sqrt{\text{HOTI}}$ and the HOTI in (\ref{eq:hamilthotibo22}).
\begin{figure*}[ht]
	\begin{centering}
		\includegraphics[width=0.95 \textwidth,height=6.5cm]{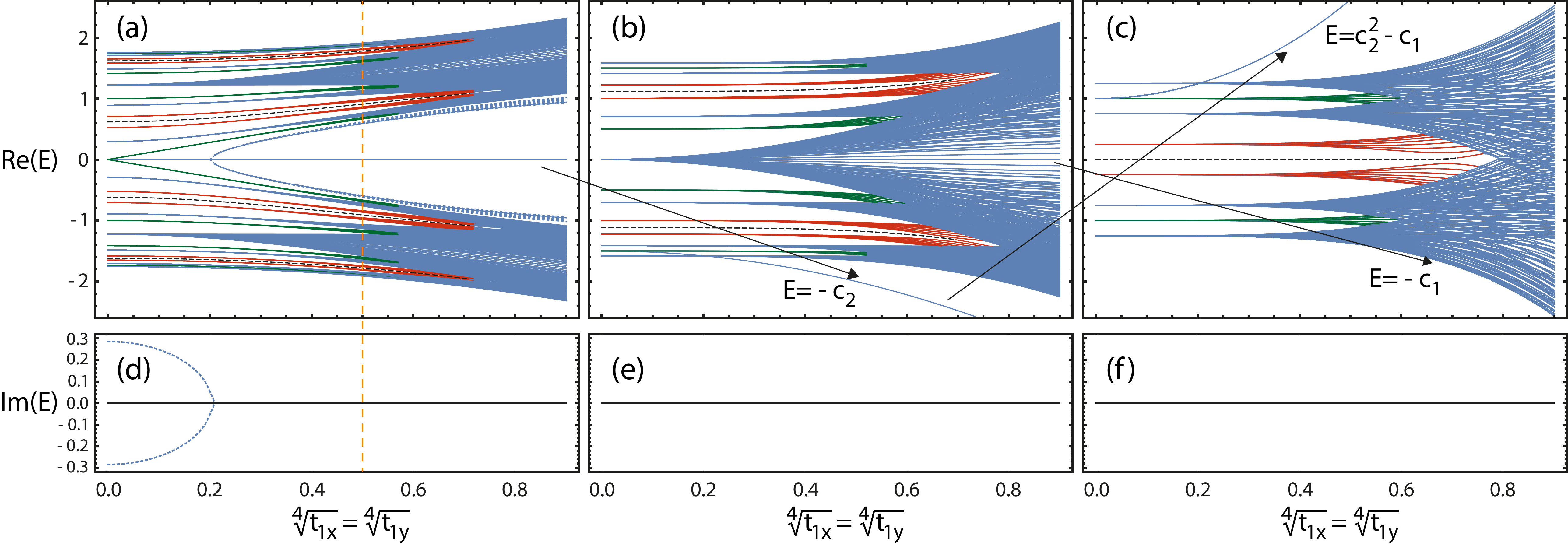} 
		\par\end{centering}
	\caption{Real part of the energy spectrum, with the same parameters and color scheme for the curves as in Fig.~\ref{fig:hotibo}(b), obtained from diagonalization of: (a) $H_{\sqrt[4]{\text{HOTI}}}$, the Hamiltonian of the open $\sqrt[4]{\text{HOTI}}$ lattice with $8\times 8$ unit cells plus 36 extra sites from incomplete unit cells (16 extra gray sites with index 35 and 36, from Fig.~\ref{fig:ucellboti4}, placed on top and 16 extra gray sites with index 16 and 28 placed at the left); (b) $H_{\sqrt[4]{\text{HOTI}}}^{2^\prime}=H_{\sqrt[4]{\text{HOTI}}}^{2}-c_2I$, with $c_2$ given in (\ref{eq:c2hotibo4}); (c) $H_{\sqrt[4]{\text{HOTI}}}^{4^\prime}=H_{\sqrt[4]{\text{HOTI}}}^{2^\prime}H_{\sqrt[4]{\text{HOTI}}}^{2^\prime}-c_1I$, with $c_1$ given below (\ref{eq:hamiltres}).
	(d)-(e) Imaginary part of the spectrum for the corresponding cases above.
    The dashed orange vertical line crossing (a) and (d) at $\sqrt[4]{t_{1x}}=0.5$ separates the non-Hermitian region to the left from the Hermitian region to the right.}
	\label{fig:espectrumhotibo4}
\end{figure*}
However, if one considers only the blue and gray sites in Fig.~\ref{fig:ucellboti4}, the self-energy is not the same at all the sites in the blue sublattice upon squaring the $\sqrt[4]{\text{HOTI}}$ bulk Hamiltonian,
\begin{eqnarray}
	\bra{j}H^2_{\sqrt[4]{\text{HOTI}}}\ket{j}&=&\sum\limits_{\substack{i=1,2, \\ \mu=x,y}}\sqrt{t_{i\mu}},\ \ \ j=1,2,3,4,
	\label{eq:sites1to4}
	\\
	\bra{5}H^2_{\sqrt[4]{\text{HOTI}}}\ket{5}&=&\bra{9}H^2_{\sqrt[4]{\text{HOTI}}}\ket{9}=2\sqrt{t_{1x}},
	\label{eq:sites5and9}
	\\
	\bra{6}H^2_{\sqrt[4]{\text{HOTI}}}\ket{6}&=&\bra{10}H^2_{\sqrt[4]{\text{HOTI}}}\ket{10}=2\sqrt{t_{2x}},
	\\
	\bra{7}H^2_{\sqrt[4]{\text{HOTI}}}\ket{7}&=&\bra{8}H^2_{\sqrt[4]{\text{HOTI}}}\ket{8}=2\sqrt{t_{1y}},
	\\
	\bra{11}H^2_{\sqrt[4]{\text{HOTI}}}\ket{11}&=&\bra{12}H^2_{\sqrt[4]{\text{HOTI}}}\ket{12}=2\sqrt{t_{2y}},
	\label{eq:sites11and12}
\end{eqnarray}
where the $\mathbf{k}=(k_x,k_y)$ dependence is implied in $H^2_{\sqrt[4]{\text{HOTI}}}$ for convenience.
In order to recover a constant energy shift $c_2$ for the entire blue sublattice, we first set it to the self-energy at sites 1 to 4 in (\ref{eq:sites1to4}),
\begin{equation}
	c_2=\sum\limits_{\substack{i=1,2, \\ \mu=x,y}}\sqrt{t_{i\mu}},
	\label{eq:c2hotibo4}
\end{equation}
and then add extra green sites in Fig.~\ref{fig:ucellboti4}, connected to sites 5 to 12 by hoppings chosen such that the difference between (\ref{eq:sites5and9})-(\ref{eq:sites11and12}) and (\ref{eq:sites1to4}) is compensated, yielding a $c_2$ self-energy term for all sites in the blue sublattice.
These new hopping terms can be readily found to be given by
\begin{equation}
	\sqrt[4]{v_{j\nu}}=\sqrt{\sum\limits_{\{i,\mu\}\neq\{j,\nu\}}\sqrt{t_{i,\mu}}-\sqrt{t_{j,\nu}}}.
\end{equation}
If we set, as in the two previous sections, $t_{2x}=1$, $t_{2y}=0.25$ and $t_{1x}=t_{1y}$, this leads to $\sqrt[4]{v_{2x}}=\sqrt{2\sqrt{t_{1x}}-0.5}$, which is purely imaginary and \textit{non-Hermitian} for $t_{1x}<0.0625$ ($\sqrt[4]{t_{1x}}<0.5$).
The non-Hermiticity of $\sqrt[4]{v_{2x}}$ in this interval can be understood by noticing that, without the extra sites, the on-site energies at sites 6 and 10 in Fig.~\ref{fig:ucellboti4}, coupled by the strong $t_{2x}$ links to the respective adjacent sites, are greater than $c_2$ when the system is squared.
In order to compensate for the difference to $c_2$, the squared extra hopping terms $\sqrt[4]{v_{2x}}$ (light blue links coupling to the extra sites 20 and 32) have to yield a \text{negative} contribution ($\sqrt{v_{2x}}<0$) in this $\sqrt[4]{t_{1x}}<0.5$ region, requiring $\sqrt[4]{v_{2x}}$ to be non-Hermitian.

The real and imaginary parts of the energy spectrum of the $\sqrt[4]{\text{HOTI}}$ are shown in Figs.~\ref{fig:espectrumhotibo4}(a) and (d), respectively, considering an open square geometry of $8\times 8$ unit cells plus 36 extra sites from incomplete unit cells (16 extra gray sites with index 35 and 36, from Fig.~\ref{fig:ucellboti4}, placed on top and 16 extra gray sites with index 16 and 28 placed at the left).
The role of the extra sites is to ensure a uniform $c_2$ on-site energy on all sites of the relevant blue sublattice upon squaring \cite{Marques2021}, such that the $\sqrt{\text{HOTI}}$ in Fig.~\ref{fig:hotibo2}(a) is recovered as one of the diagonal blocks of the squared Hamiltonian \footnote{Note that adding extra sites from the larger sublattice has a trivial effect on the energy spectrum, namely the addition of the same amount of zero-energy states due to sublattice imbalance.}.
The dashed orange vertical line separates the non-Hermitian region to the left from the Hermitian region to the right.
Within the non-Hermitian region, two additional subregions can be distinguished: i) a parity-time ($\mathcal{PT}$) symmetry breaking \cite{Bender2007} subregion below the exceptional point \cite{Bergholtz2021} at $\sqrt[4]{t_{1x}}\approx 0.2$, characterized by the appearance of a set of eigenstates with imaginary energies, given by the dashed blue curves in Fig.~\ref{fig:espectrumhotibo4}(d), and ii) a $\mathcal{PT}$-symmetry preserving subregion for $0.2\lesssim\sqrt[4]{t_{1x}}<0.5$, characterized by a purely real spectrum [the imaginary energies evolve into the symmetric blue dashed curves of bulk states bifurcating from $E=0$ at $\sqrt[4]{t_{1x}}\approx 0.2$ in Fig.~\ref{fig:espectrumhotibo4}(a)].
Four bands or corner states, given by the dashed black curves, are present in Fig.~\ref{fig:espectrumhotibo4}(a), together with a plethora of 1D vertical (green curves) and horizontal (red curves) bands of edge states that are driven into the bulk blue bands in the Hermitian region.

Squaring the $\sqrt[4]{\text{HOTI}}$ bulk Hamiltonian leads to
\begin{eqnarray}
H^2_{\sqrt[4]{\text{HOTI}}}(\mathbf{k})&=&
\begin{pmatrix}
H_{\sqrt{\text{par},2}}&0
\\
0&H_{\sqrt{\text{res},2}}
\end{pmatrix},
\label{eq:hamilthotibo42}
\\
H_{\sqrt{\text{par},2}}&=&c_2 I_{12}+H_{\sqrt{\text{HOTI}}},
\label{eq:hamiltpar2}
\\
H_{\sqrt{\text{res},2}}&=&c_2 I_{24}+H_{\sqrt{\text{res},2^\prime}},
\label{eq:hamiltres2}
\end{eqnarray}
where $c_2$ is given in (\ref{eq:c2hotibo4}), $H_{\sqrt{\text{HOTI}}}$ is the Hamiltonian of the $\sqrt{\text{HOTI}}$ given in (\ref{eq:hamilthotibo2}), while $H_{\sqrt{\text{res},2^\prime}}$ is the Hamiltonian of a residual lattice involving the sites in the gray and green sublattice of Fig.~\ref{fig:ucellboti4}, which has the same spectrum as the $\sqrt{\text{HOTI}}$ apart from the extra 12 bands with $E=-c_2$ [zero-energy bands, if the constant shift in (\ref{eq:hamilthotibo42}) is included] coming from the sublattice imbalance.

The real and imaginary parts of the squared energy spectrum, with a global downshift of $c_2$ given in (\ref{eq:c2hotibo4}), of the open  $\sqrt[4]{\text{HOTI}}$ lattice of Figs.~\ref{fig:espectrumhotibo4}(a) and (d) is shown in Figs.~\ref{fig:espectrumhotibo4}(b) and (e), respectively.
The spectrum is purely real and coincides with that of the $\sqrt{\text{HOTI}}$ lattice in Fig.~\ref{fig:hotibo2}(b), as expected, apart from the degenerate lowest energy curve with $E=-c_2$.
Interestingly, the imaginary part of the spectrum vanishes everywhere in Fig.~\ref{fig:espectrumhotibo4}(e), even though the the residual block $H_{\sqrt{\text{res},2^\prime}}$ in (\ref{eq:hamilthotibo42}) remains non-Hermitian for $\sqrt[4]{t_{1x}}<0.5$.
It is easy to see that, in this region, upon squaring the lattice in Fig.~\ref{fig:ucellboti4} one obtains a non-Hermitian $i$ phase factor in the hopping connecting, e.g., sites 15 and 20, picked up along the light blue $\sqrt[4]{v_{2x}}$ (recall that $H_{\sqrt{\text{res},2^\prime}}$ comprises the sites in the gray and green sublattice).
On the other hand, we know that the relevant blue sublattice remains Hermitian upon squaring, since the non-Hermitian $\sqrt[4]{v_{2x}}$ couplings, for $\sqrt[4]{t_{1x}}<0.5$, translate by design as an extra real and negative on-site contribution at sites 6 and 10 in Fig.~\ref{fig:ucellboti4}.
Since, as shown in Appendix~\ref{app:degenspectrum}, $H_{\sqrt{\text{par},2}}$ shares the same finite energy spectrum as $H_{\sqrt{\text{res},2}}$, with the extra zero-energy states coming from sublattice imbalance, it follows that $H_{\sqrt{\text{res},2}}$ displays a purely real spectrum even within the non-Hermitian region.
This constitutes a new mechanism that should be distinguished from that reported in recent studies of square-root topology in non-Hermitian models \cite{Ezawa2020,Ke2020,Lin2021}, where the same  $\mathcal{PT}$-symmetry broken region with complex energy spectrum is found in both the original and squared models.

Finally, by squaring the energy downshifted versions of $H^2_{\sqrt[4]{\text{HOTI}}}$ in (\ref{eq:hamilthotibo42}), and subsequently downshifting by $c_1=t_{1x}+t_{2x}+t_{1y}+t_{2y}$ (the resulting on-site energy at the relevant HOTI sublattice),  one arrives at
\begin{eqnarray}
	H^4_{\sqrt[4]{\text{HOTI}}^\prime}(\mathbf{k})&:=&\big(H^2_{\sqrt[4]{\text{HOTI}}}-c_2I_{36}\big)^2-c_1I_{36} \nonumber
	\\
	&=&
	\begin{pmatrix}
	H_{\text{par}}&0&0
	\\
	0&H_{\text{res}}&0
	\\
	0&0&H_{\text{res},2}
	\end{pmatrix}-c_1I_{36}, \nonumber
	\\
	&=&\begin{pmatrix}
	H_{\text{HOTI}}&0&0
	\\
	0&H_{\text{res}^\prime}&0
	\\
	0&0&H_{\text{res,2}^\prime}
	\end{pmatrix},
	\label{eq:hamilthotibo44}
	\\
	H_{\text{res},2}&=&c_1I_{24}+H_{\text{res,2}^\prime},
\end{eqnarray}
with $H_{\text{par}}$ and $H_{\text{res}}$ given, respectively, in (\ref{eq:hamiltpar}) and (\ref{eq:hamiltres}).
The real and imaginary parts of the energy spectrum of $H^4_{\sqrt[4]{\text{HOTI}}^\prime}$, coming from the open  $\sqrt[4]{\text{HOTI}}$ lattice of Figs.~\ref{fig:espectrumhotibo4}(a) and (d), is shown in Figs.~\ref{fig:espectrumhotibo4}(c) and (f), respectively.
Even though the decoupled lattice corresponding to $H_{\text{res,2}}=H_{\sqrt{\text{res},2^\prime}}^2$ still has non-Hermitian terms for $\sqrt[4]{t_{1x}}<0.5$, the energy spectrum is again purely real, which can be demonstrated by applying the same reasoning as above for the squared spectrum in Fig.~\ref{fig:espectrumhotibo4}(b).
As indicated by the arrows, the degenerate zero-energy curve in Fig.~\ref{fig:espectrumhotibo4}(a), originated from sublattice imbalance in $H_{\sqrt[4]{\text{HOTI}}}$, becomes the $E=-c_2$ curve in Fig.~\ref{fig:espectrumhotibo4}(b) and the $E=c_2^2-c_1$ curve in Fig.~\ref{fig:espectrumhotibo4}(c).
The degenerate zero-energy curve in Fig.~\ref{fig:espectrumhotibo4}(b), originated both from the sublattice imbalance within $H_{\sqrt{\text{HOTI}}}$ in (\ref{eq:hamiltpar2}) and from the degenerate finite energy spectrum of $H_{\sqrt{\text{par,2}}}$ in (\ref{eq:hamiltpar2}) and $H_{\sqrt{\text{res,2}}}$ in (\ref{eq:hamiltres2}), becomes the $E=-c_1$ curve in Fig.~\ref{fig:espectrumhotibo4}(c), which delimits the lowest bulk continuum from below.

Aside from these extra bands that are present due to successive sublattice imbalances and degeneracies, the energy spectrum in Fig.~\ref{fig:espectrumhotibo4}(c) corresponds to a four-fold degenerate spectrum of the HOTI lattice in Fig.~\ref{fig:hotibo}(b), with one degeneracy coming from $H_{\text{res}^\prime}$ and two others from $H_{\text{res,2}^\prime}$ in (\ref{eq:hamilthotibo44}).
In particular, this four-fold degeneracy affects the 1D edge states, both the vertical ones at the bottom edge and the horizontal ones at the right edge (see their location in Fig.~\ref{fig:squaredhotibo2}),  and the 0D higher-order states at the right bottom corner.
Note, however, that only one these states in each fourfold-degenerate group is topological in nature, coming from $H_{\text{HOTI}}$, while the other three are impurity states, one coming from $H_{\text{res}^\prime}$ and two from $H_{\text{res,2}^\prime}$.
Generalizing what was shown in the previous section for the $\sqrt{\text{HOTI}}$ lattice, we can say that each boundary mode of the starting $\sqrt[4]{\text{HOTI}}$ lattice in Fig.~\ref{fig:espectrumhotibo4}(a) can be written as a linear combination of the corresponding four degenerate states in Fig.~\ref{fig:espectrumhotibo4}(c), all with a quarter of their weight on the topological state within the group.
As such, both the weak edge and the higher-order corner states of the $\sqrt[4]{\text{HOTI}}$ lattice can be said to be a ``quarter'' topological and ``three quarters'' impurity, with their topological features directly inherited from the $\ket{\psi^\text{HOTI}_j}$ component.
Thus, the $\sqrt[4]{\text{HOTI}}$ model with the unit cell of Fig.~\ref{fig:ucellboti4} constitutes an example of a quartic-root (or 4-root) weak and higher-order topological insulator.

It is clear at this point that this method can be repeated to find the 8-root HOTI, 16-root HOTI, etc.
As the root-degree increases so will the size of the unit cell.
Let us summarize the process for the construction of the $2^n$-root HOTI, with $n\in\mathbb{N}$: 
\\
\\
(i) One starts by identifying all sites in the $2^{n-1}$-root model as the relevant sublattice of the $2^n$-root model (which we depict in blue as exemplified in Fig.~\ref{fig:squaredhotibo2}, where the blue sublattice in the $\sqrt{\text{HOTI}}$ corresponds to the sites of the HOTI). 
\\
\\
(ii) One subdivides the lattice \cite{Ezawa2020} by introducing an extra site (depicted here in gray, as shown in for the $\sqrt{\text{HOTI}}$ of Fig.~\ref{fig:hotibo2}) in the middle of every hopping link such that it splits in two, each connecting the new site to one of the two adjacent blue sites. 
Together with the assumption that the original HOTI has no mass terms, that is, no modulation on its on-site energies, this step ensures that the $\sqrt[2^{n}]{\text{HOTI}}$ under construction is bipartite, such that its squared Hamiltonian can be written in a block diagonal form.
\\
\\
(iii) The phases and magnitudes of the new hopping parameters are renormalized as $\sqrt[2^{n-1}]{t_{ij}} e^{i\phi_{ij}}\to \sqrt[2^{n}]{t_{ij}} e^{i\phi_{ij}/2}$, which keeps the same flux pattern of the $\sqrt[2^{n-1}]{\text{HOTI}}$ (no phases have been introduced so far, but they will appear in some of the models studied below). 
\\
\\
(iv) One checks the squared on-sites energies of the different types of sites in the blue sublattice, based on the number and magnitudes of their hopping connections. If there is at least one discrepancy, one sets the squared on-site energy of one of the types of blue sites as the uniform level and  compensates for the difference in the other types by including extra sites (depicted here in green, as in Fig.~\ref{fig:ucellboti4}) connected to them by appropriately tuned hopping terms. 
Note that within the global constraint of having to compensate for a fixed value of the energy difference, the number of extra sites included and the amplitudes of the new hopping terms are degrees of freedom of the method.
This step is skipped whenever it is not necessary, that is, whenever there is already a uniform squared on-site energy for all blue sites.
Typically we set the type of blue sites with highest squared on-site energy for the entire parameter range, if there is one, as the constant energy level, in order to avoid the inclusion of purely imaginary non-Hermitian hoppings to the other blue sites (to shift down their higher squared on-site energy).
However, even if non-Hermitian terms are to be included, the successive squared spectra remain purely real, due the shared finite spectrum of the parent block containing the lower-root degree Hamiltonian (which is real by default) and the residual block at any given level after each squaring of the Hamiltonian (see Appendix~\ref{app:degenspectrum}).
\\
\\
(v) Under OBC and for $n>1$ (above the square-root model), blue sites along specific edges and corners have fewer connections than their bulk counterparts, which leads to a negative squared on-site energy offset at these sites, relative to the bulk ones. 
Extra sites from incomplete unit cells, placed along the directions with a connectivity deficit, must be included in the lattice. 
For the open $\sqrt[2^{n}]{\text{HOTI}}$ lattice, with $n>1$, the number of included extra sites is such that it must guarantee that, after $n-1$ successive squaring operations to the Hamiltonian followed by a downshift of the $c_{2^{n-1}}$ term of the respective level, one arrives at the $\sqrt{\text{HOTI}}$ as one of the diagonal blocks with uniform on-site energy $c_2$.
\\
\\
We point out that, even though the method outlined above for constructing $2^n$-root models is not unique, the required criteria underlying it are, namely, (i) bipartition of the higher root-degree lattice, (ii) ensuring the same flux pattern upon squaring the model, and (iii) ensuring a uniform squared on-site energy at the relevant blue sublattice. As an example, we refer the reader to [\onlinecite{Wu2021}], where a different approach was followed for the construction of the square-root version of the 2D SSH model with NNN hoppings which, however, still obeys criteria (i)-(iii).

\section{Quartic-root Chern Insulator}
\label{sec:sr4ci}

Here, we apply the method outlined at the end of the last section in the construction of the 4-root version of a well-known Chern insulator (CI), namely, the Haldane model \cite{Haldane1988}, with the change in notation HOTI$\to$CI.

\subsection{Original CI model}
\begin{figure*}[ht]
	\begin{centering}
		\includegraphics[width=0.95 \textwidth,height=9.5cm]{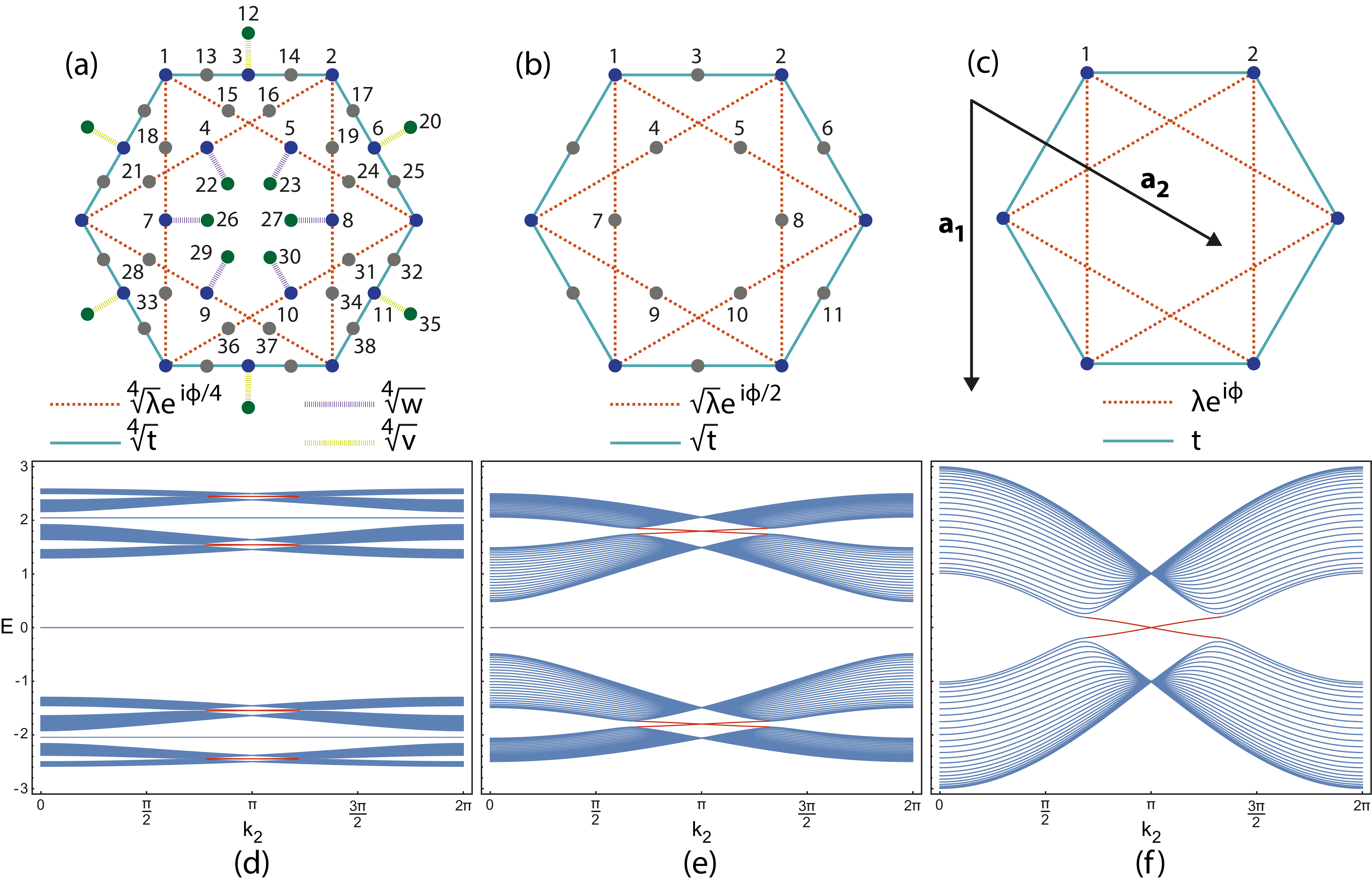} 
		\par\end{centering}
	\caption{Plaquette of the (a) quartic-root, (b) square-root, and (c) original version of the CI, corresponding to the Haldane model without a mass term. Numbered sites indicate the respective unit cells.
	Energy spectrum as a function of $k_2$, with $(t,\lambda,\phi)=(1,\frac{0.2}{3\sqrt{3}},\frac{\pi}{2})$ of: (d) $H_{\sqrt[4]{\text{CI}}}$, (e) $H_{\sqrt[4]{\text{CI}}}^{2^\prime}=H_{\sqrt[4]{\text{CI}}}^{2}-c_2I$, with $c_2$ given below (\ref{eq:vhopping}), and (f) $H_{\sqrt[4]{\text{CI}}}^{4^\prime}=H_{\sqrt[4]{\text{CI}}}^{2^\prime}H_{\sqrt[4]{\text{CI}}}^{2^\prime}-c_1I$, with $c_1$ given below (\ref{eq:hamiltressrci}) and $H_{\sqrt[4]{\text{CI}}}$ the Hamiltonian of the $\sqrt[4]{\text{HOTI}}$ lattice open along $\mathbf{a_1}$ and periodic along $\mathbf{a_2}$, with 24 extra sites in incomplete unit cells above the top edge in $\mathbf{a_1}$, given by the 6 blue sites $\{4,7,8,9,10,11\}$ plus the 18 gray and green sites connected to them in (a).
    Blue (red) bands correspond to bulk (edge) bands.}
	\label{fig:espectrumci4}
\end{figure*}

We take our original CI to be the Haldane model without the mass term (zero on-site energy for both sites within a unit cell) depicted in Fig.~\ref{fig:espectrumci4}(c).
The complex next-nearest-neighbor (NNN) hoppings break time-reversal symmetry, lifting the degeneracy at the Dirac points in the bulk spectrum and opening a gap between the two bands given by
\begin{equation}
\Delta E=6\sqrt{3}\lambda\sin\phi .
\end{equation}
We consider a cylindrical geometry with zigzag edges, with an open system along $\mathbf{a_1}$ and periodic along $\mathbf{a_2}$.
The momentum $k_2$ along the latter direction is a good quantum number and the energy spectrum of the Haldane model, our original CI, has the form of Fig.~\ref{fig:espectrumci4}(f) for $t=1$ as the energy unit, $\lambda=\frac{0.2}{3\sqrt{3}}$ and $\phi=\frac{\pi}{2}$.
A chiral pair of red edge bands crossing at zero-energy appears at the gap between the two bulk bands, which can be related to the non-trivial Chern number $C=1$ of the lower bulk band.

\subsection{$\sqrt{\text{CI}}$ model}
The $\sqrt{\text{CI}}$ has a unit cell with 11 sites, as shown in Fig.~\ref{fig:espectrumci4}(b).
Its bulk Hamiltonian in the ordered $\{\ket{j,\mathbf{k}}\}$ basis, with $j=1,2,\dots,11$, with $\mathbf{k}=(k_1,k_2)$ and assuming that $\mathbf{a_1}=a_1\hat{e}_x$ and $\mathbf{a_2}=a_2\hat{e}_y$ are orthogonalized by square-shaping the geometry of the model, and setting also $a_1=a_2=1$, reads as
\begin{widetext}
	\begin{eqnarray}
	H_{\sqrt{\text{CI}}}(\mathbf{k})&=&
	\begin{pmatrix}
	0&h_{\sqrt{\text{CI}}}^\dagger
	\\
	h_{\sqrt{\text{CI}}}&0
	\end{pmatrix},
	\label{eq:hamiltci2}
	\\
	h_{\sqrt{\text{CI}}}&=&
	\begin{pmatrix}
	\sqrt{t}&\sqrt{t}
	\\
	0&\sqrt{\lambda}\big(e^{-i\frac{\phi}{2}}+e^{-i(k_2-k_1-\frac{\phi}{2})}\big)
	\\
	\sqrt{\lambda}\big(e^{i\frac{\phi}{2}}+e^{-i(\frac{\phi}{2}-k_2)}\big)&0
	\\
	\sqrt{t}e^{ik_2}&\sqrt{t}
	\\
	\sqrt{\lambda}\big(e^{-i\frac{\phi}{2}}+e^{i(\frac{\phi}{2}+k_1)}\big)&0
	\\
	0&\sqrt{\lambda}\big(e^{i\frac{\phi}{2}}+e^{-i(\frac{\phi}{2}-k_1)}\big)
	\\
	0&\sqrt{\lambda}\big(e^{i(\frac{\phi}{2}+k_1)}+e^{-i(\frac{\phi}{2}+k_2-k_1)}\big)
	\\
	\sqrt{\lambda}\big(e^{-i(\frac{\phi}{2}-k_1)}+e^{i(\frac{\phi}{2}+k_2)}\big)&0
	\\
	\sqrt{t}e^{ik_2}&\sqrt{t}e^{ik_1}
	\end{pmatrix},
	\end{eqnarray}
\end{widetext}
yielding, when squared,
\begin{equation}
H_{\sqrt{\text{CI}}}^2(\mathbf{k})=c_1I_{11}+
	\begin{pmatrix}
H_{\text{CI}}(\mathbf{k})&0
\\
0&H_{\text{res}^\prime}(\mathbf{k})
\label{eq:hamiltressrci}
\end{pmatrix},
\end{equation}
where $c_1=3t+6\lambda$ is a constant energy shift, $H_{\text{CI}}(\mathbf{k})=h_{\sqrt{\text{CI}}}^\dagger h_{\sqrt{\text{CI}}}-c_1I_2$ is the two-band bulk Hamiltonian of the CI model in Fig.~\ref{fig:espectrumci4}(c), and $H_{\text{res}^\prime}(\mathbf{k})=h_{\sqrt{\text{CI}}}h_{\sqrt{\text{CI}}}^\dagger-c_1I_9$ is the nine-band bulk Hamiltonian of a residual Hamiltonian built on the gray sublattice of Fig.~\ref{fig:espectrumci4}(b).
Notice that step (iv), regarding the inclusion of extra sites, was skipped for the construction of the $\sqrt{\text{CI}}$, since both blue sites in the unit cell of Fig.~\ref{fig:espectrumci4}(b) already have the same squared on-site energy $c_1$.
Also, the $\sqrt{\text{CI}}$ is bipartite and can be subdivided in a blue and a gray sublattice, even though the CI itself is not bipartite,

The energy spectrum of the $\sqrt{\text{CI}}$ for the same ribbon geometry as before, with the same parameters as Fig.~\ref{fig:espectrumci4}(f), has the form of Fig.~\ref{fig:espectrumci4}(e).
Two symmetric pairs of edge bands are found, with the crossing between the bands of each pair occurring at $k_2=\pi$ and $E=\pm\sqrt{c_1}\approx\pm 1.8$.
However, it should be emphasized that we included the set of six extra sites $\{4,7,8,9,10,11\}$ in incomplete unit cells above the top edge already at this $\sqrt{\text{CI}}$ stage, contrary to what was stated in step (v), which only applies above square-root topology.
These extra sites were included to enable a direct comparison between Fig.~\ref{fig:espectrumci4}(e) and Fig.~3(d) in [\onlinecite{Ezawa2020}], where they are implicitly included.
Without them, only one edge band would be found for positive energy and a symmetric one with negative energy, with both lying at the lower edge.
The sevenfold-degenerate zero-energy band is a consequence of the sublattice imbalance.
If one squares this spectrum and applies a general energy downshift of $c_1$ one arrives, apart form the extra degenerate bands, to the spectrum of Fig.~\ref{fig:espectrumci4}(f) with double degeneracy for all bands, including the red edge bands, summing to a total of four, with two topological edge bands coming from the CI lattice and two impurity edge bands coming from the residual lattice.

\subsection{$\sqrt[4]{\text{CI}}$ model}

Starting by identifying all the sites in the $\sqrt{\text{CI}}$ in Fig.~\ref{fig:espectrumci4}(b) as the blue sublattice of the $\sqrt[4]{\text{CI}}$, we can then apply all the steps towards the construction of this model, which leads to the $\sqrt[4]{\text{CI}}$ depicted in Fig.~\ref{fig:espectrumci4}(a).
The size of the unit cell is considerably enlarged to 38 sites, including nine extra green sites connected to the blue sublattice by two newly defined hopping terms,
\begin{eqnarray}
	\sqrt[4]{w}&=&\sqrt{3\sqrt{t}+4\sqrt{\lambda}},
	\\
	\sqrt[4]{v}&=&\sqrt{\sqrt{t}+6\sqrt{\lambda}},
	\label{eq:vhopping}
\end{eqnarray} 
introduced in order to ensure a uniform squared on-site energy for all sites in the blue sublattice, given by $c_2=3\sqrt{t}+6\sqrt{\lambda}$.

In Fig.~\ref{fig:espectrumci4}(d), we show the energy spectrum of the $\sqrt[4]{\text{CI}}$ in Fig.~\ref{fig:espectrumci4}(a), obtained from diagonalization of $H_{\sqrt[4]{\text{CI}}}$ in a ribbon geometry with 24 extra sites in incomplete unit cells above the top edge, given by the 6 blue sites $\{4,7,8,9,10,11\}$ plus the 18 gray and green sites connected to them, for a parameter set $(t,\lambda,\phi)=(1,\frac{0.2}{3\sqrt{3}},\frac{\pi}{2})$.
Four edge bands are found, with the crossing between the bands of each pair occurring at $k_2=\pi$ and $E=\pm\sqrt{c_2\pm\sqrt{c_1}}$.
The zero-energy 16-fold-degenerate bulk band appears due to sublattice imbalance.

In Fig.~\ref{fig:espectrumci4}(e), we show the energy spectrum obtained from diagonalization of $H^{2^\prime}_{\sqrt[4]{\text{CI}}}=H^2_{\sqrt[4]{\text{CI}}}-c_2I=\text{diag}(H_{\sqrt{\text{CI}}},H_{\sqrt{\text{Res},2^\prime}})$, with $I$ the identity matrix of $\dim(I)=\dim (H_{\sqrt[4]{\text{CI}}})$ and $H_{\sqrt{\text{Res},2^\prime}}$ the square-root of the residual lattice built on the gray plus green sublattice of Fig.~\ref{fig:espectrumci4}(a), with impurity sites along the top and bottom edges.
The global spectrum is a superposition of two identical spectra, one corresponding to each of the diagonal blocks.
In particular, each red edge band is therefore doubly-degenerate, with one stemming from the $\sqrt{\text{CI}}$ and the other from the square-root residual block.
The zero-energy states in Fig.~\ref{fig:espectrumci4}(d) become the $E=-c_2$ states of $H_{\sqrt{\text{Res},2^\prime}}$, which fall below the energy range of Fig.~\ref{fig:espectrumci4}(e).

The energy spectrum obtained from diagonalization of
\begin{eqnarray}
H^{4^\prime}_{\sqrt[4]{\text{CI}}}&=&H^{2^\prime}_{\sqrt[4]{\text{CI}}}H^{2^\prime}_{\sqrt[4]{\text{CI}}}-c_1I \nonumber
\\
&=&
\begin{pmatrix}
H_{\text{CI}}&&
\\
&H_{\text{Res}^\prime}&
\\
&&H_{\text{Res},2^\prime}
\end{pmatrix}
\end{eqnarray} 
is shown in Fig.~\ref{fig:espectrumci4}(f), apart from the extra states with energy $E=-c_1$ ($E=c_2^2-c_1$) which fall below (above) the energy range.
The global visible spectrum consists of a superposition of four identical spectra, one coming from $H_{\text{CI}}$, one from $H_{\text{Res}^\prime}$, and two from $H_{\text{Res},2^\prime}$.
For each fourfold-degenerate red edge band, only one is topological in nature, stemming from the Haldane lattice, our original CI, while the other three are impurity bands stemming from the two successive residual chains.
Furthermore, each edge state in Fig.~\ref{fig:espectrumci4}(d) can be written as a linear combination of the corresponding four degenerate states in Fig.~\ref{fig:espectrumci4}(f) and shown to have exactly one quarter of its weight on the topological state of this set, defining the model in Fig.~\ref{fig:espectrumci4}(a) as a quartic-root CI.

A subsequent application of the method outlined at the end of Sec.~\ref{sec:qrasym2dssh} on the $\sqrt[4]{\text{CI}}$ would yield the $\sqrt[8]{\text{CI}}$, from which its even higher root-degree versions $\sqrt[2^n]{\text{CI}}$, with $n>3$, could be successively found.
This method could also be applied to find the higher-root degree versions of a modulated Haldane model recently studied \cite{Wang2021} (provided the mass term is set to zero), which was shown to have higher-order boundary modes for a region in parameter space.

\section{Quartic-root breathing kagome lattice}
\label{sec:sr4hoti}

\begin{figure*}[ht]
	\begin{centering}
		\includegraphics[width=0.85 \textwidth,height=8.5cm]{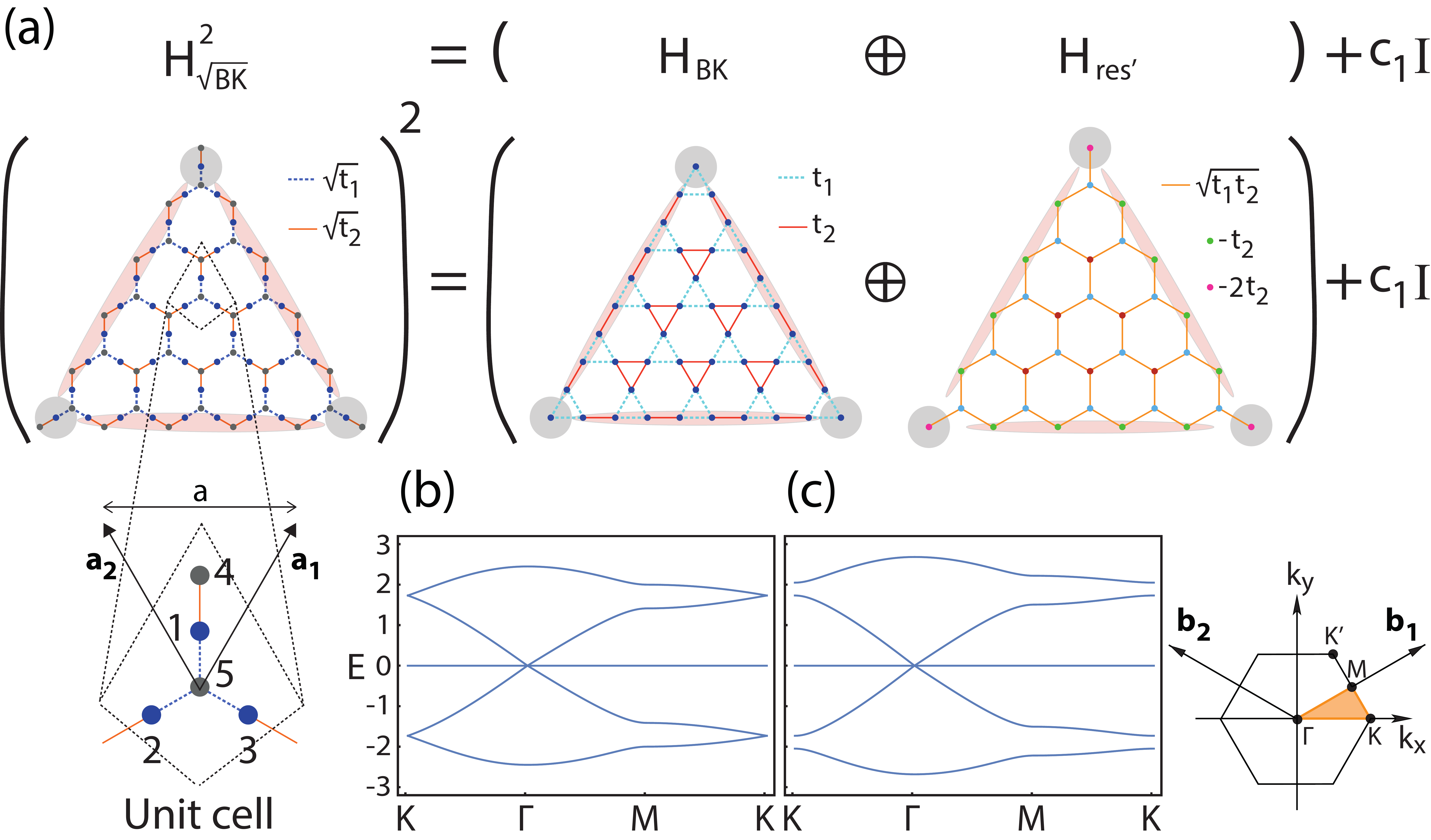} 
		\par\end{centering}
	\caption{(a) Schematic illustration of the result of squaring the $\sqrt{\text{BK}}$ triangular lattice (and the respective Hamiltonian above), with the unit cell shown at the lower left. Up to a constant overall energy shift $c_1$, the squared model is composed of two decoupled lattices, one yielding the BK in the blue sublattice and the other yielding the residual model in the gray sublattice with edge (corner) impurity sites with an on-site energy offset of $-t_2$ ($-2t_2$), in relation to the red bulk sites. 
	Shaded red (gray) regions indicate the localization of the edge (corner) states.	
	(b), (c) Energy spectrum of the bulk $\sqrt{\text{BK}}$ model, in units of $\sqrt{t_1}=1$, along the high-symmetry lines delimiting the orange region of the Brillouin zone depicted at the right for (b) $\sqrt{t_2}=\sqrt{t_1}$ and (c) $\sqrt{t_2}=1.2$. High-symmetry points of the Brillouin zone: $\Gamma=(0,0)$, $K=(\frac{4\pi}{3},0)$, $K^\prime=(\frac{2\pi}{3},\frac{2\pi}{\sqrt{3}})$ and $M=(\pi,\frac{\pi}{\sqrt{3}})$.}
	\label{fig:srkagome}
\end{figure*}
In this section, we will derive the quartic-root version of the breathing kagome (BK) lattice \cite{Hatsugai2011,Ezawa2018,Kunst2018,Kempkes2019,Wakao2020,Proctor2021,Zhong2021}.
We will follow the method outlined at the end of Sec.~\ref{sec:qrasym2dssh}, with the change in notation HOTI$\to$BK, when finding the $\sqrt[4]{\text{BK}}$ model, while the $\sqrt{\text{BK}}$ model was taken directly from existing literature, as explained in more detail in Sec.~\ref{sec:sr4hoti}-B.

\subsection{Original BK model}

\begin{figure*}[ht]
	\begin{centering}
		\includegraphics[width=0.95 \textwidth,height=5.cm]{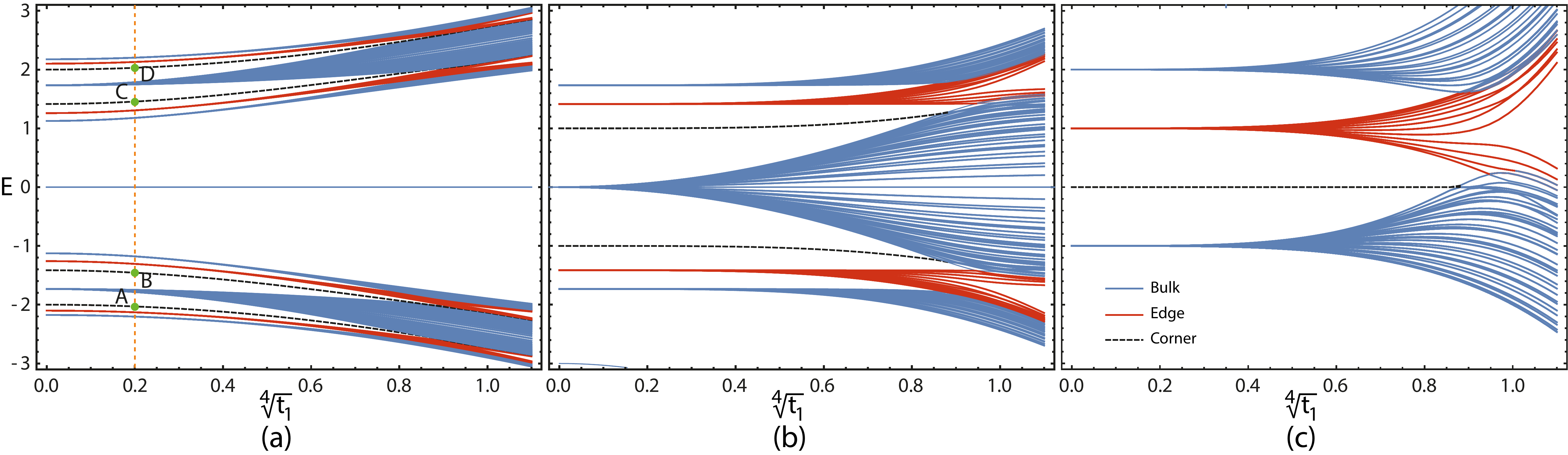} 
		\par\end{centering}
	\caption{Energy spectrum, in units of $t_2=1$, as a function of a linear variation of $\sqrt[4]{t_{1}}$, obtained from diagonalization of: (a) $H_{\sqrt[4]{\text{BK}}}$, the Hamiltonian of the $\sqrt[4]{\text{BK}}$ triangular lattice with the unit cell of Fig.~\ref{fig:highsymbk4}(a); (b) $H_{\sqrt[4]{\text{BK}}}^{2^\prime}=H_{\sqrt[4]{\text{BK}}}^{2}-c_2I$, with $c_2=3(\sqrt{t_1}+\sqrt{t_2})$; (c) $H_{\sqrt[4]{\text{BK}}}^{4^\prime}=H_{\sqrt[4]{\text{BK}}}^{2^\prime}H_{\sqrt[4]{\text{BK}}}^{2^\prime}-c_1I$, with $c_1=t_1+t_2$.
	The different types of states are indicated in (c).
    Green labeled points along the dashed orange vertical line in (a) correspond to the respective states depicted in Fig.~\ref{fig:highsymbk4}(d).}
	\label{fig:espectrumbk4}
\end{figure*}

Let us consider the BK lattice, which is a kagome lattice with alternating hopping terms as depicted in the middle of Fig.~\ref{fig:srkagome}(a).
This model has been shown to host both conventional (edge) and higher-order (corner) boundary modes for $t_1<t_2$ located, respectively, around the red and gray shaded regions.
The energy spectrum of this model for a triangular configuration, as a function of a linear variation of $\sqrt[4]{t_1}$, has the form of Fig.~\ref{fig:espectrumbk4}(c).
The spectrum displays two blue continua of bulk states related, in the corresponding bulk spectrum under PBC, to a single bulk band for the top one and to two bulk bands for the bottom one (one being a flat band following the lowest blue curve), two red continua of edge states, with one visible in the gap between the bulk continua around $E=1$ and the other one buried within the lower bulk continuum, and a three-fold degenerate zero-energy corner band (dashed black curve) in the $\sqrt[4]{t_1}<1$ region, but gapped only in the $\sqrt[4]{t_1}\lesssim 0.9$ in Fig.~\ref{fig:espectrumbk4}(c). 
The corner states disappear above the simultaneous bulk and edge band gap closing point $t_1=t_2$ (remember that the lower continuum of edge states is buried within the lower continuum of bulk states).

Some studies have identified these corner modes with topological higher-order states, proposed to be protected either by (i) a generalized chiral symmetry \cite{Kempkes2019,Ni2019} allowing for perturbations on the hopping terms between sites in different sublattices, or (ii) a combination of the three-fold rotation symmetry ($C_3$-symmetry) with its three mirror symmetries \cite{Ezawa2018}.
The polarization vector $\mathbf{P}=p_1\mathbf{a_1}+p_2\mathbf{a_2}$ is used as the  $\mathbb{Z}_3$ topological index of the system \cite{Fang2012}, since $p_1=p_2=0,\frac{e}{3},\frac{2e}{3}\ \mod e$.
However, these conclusions have been challenged by Miert and Ortix in a recent paper \cite{Miert2020}, where the authors considered a specific set of perturbations that respected the symmetries in (i) and (ii) and showed how, nevertheless, the corner states can be gapped out without closing either the bulk or edge bands.
Consequently, the corner modes were characterized as conventional higher-order states without topological protection.
For the sake of completeness, as well as for comparison with related works, we will still calculate the quantized polarization of the $\sqrt{\text{BK}}$ and $\sqrt[4]{\text{BK}}$ lattices analyzed next. 
On the other hand, another recent study \cite{Jung2021} relates the zero-energy corner states to the topology of the lower edge band, which displays a non-trivial Zak phase.
Due to this ongoing debate, we refrain from topological considerations when discussing the different $\sqrt[2^n]{\text{BK}}$ models below.

\subsection{$\sqrt{\text{BK}}$ model}

The $\sqrt{\text{BK}}$ model, shown in a triangular configuration at the left of Fig.~\ref{fig:srkagome}(a), describes a decorated honeycomb lattice, also called super-honeycomb \cite{Shima1993,Aoki1996,Zhong2017}, edge-centered honeycomb \cite{Lan2012} or honeycomb-kagome lattice \cite{Lu2017,Barreteau2017}, which has been experimentally realized in photonic lattices \cite{Yan2020a,Maimati2020} and predicted to be the ground state configuration of several monolayer materials \cite{Song2016,Wang2017,Ji2017,Wang2017b,Ding2017,Liu2017b,Pan2018,Song2019,Song2019b,Zhang2021}.
Note that this $\sqrt{\text{BK}}$ model was not derived from our proposed scheme outlined at the end of Sec.~\ref{sec:qrasym2dssh}.
Instead, we chose it from comparison with existing literature \cite{Mizoguchi2020} and for allowing it also to serve, in the limit of constant hopping terms, as the square-root version of the topological semimetal analyzed in the next section.
However, one could have chosen to follow our method to derive the $\sqrt{\text{BK}}$ lattice, yielding a subdivided version of the BK lattice in the middle of Fig.~\ref{fig:srkagome}(a), which would not change the fundamental properties of the system (e.g., it would have the same energy gaps, the same edge and corner energy levels, etc.).

The bulk energy spectrum of the $\sqrt{\text{BK}}$ model along the high-symmetry lines is shown in Fig.~\ref{fig:srkagome}(b) for $\sqrt{t_1}=\sqrt{t_2}$, where a semimetallic behavior is found for a Fermi level set at the band touching at the $K$ point between the top or bottom pairs of bands, and in Fig.~\ref{fig:srkagome}(c) for $\sqrt{t_1}\neq\sqrt{t_2}$, where the degeneracies are lifted at this point and a gap opens at symmetric energies \cite{Mizoguchi2020}.
In both cases a spin-1 Dirac cone is found centered at zero-energy around the $\Gamma$ point, reminiscent of the low-energy behavior of the Lieb \cite{Flannigan2021} and other kagome-type\cite{Mizoguchi2021} lattices.

The $\sqrt{\text{BK}}$ model has a three-fold rotation symmetry about the $z$ axis centered at site 5 defined, in the ordered $\{\ket{j,\mathbf{k}}\}$ basis, with $j=1,2,\dots,5$, as
\begin{eqnarray}
C_3:\ &R_3(\mathbf{k})&H_{\sqrt{\text{BK}}}(\mathbf{k})R_3^{-1}(\mathbf{k})=H_{\sqrt{\text{BK}}}\bigg(R\Big(\frac{2\pi}{3}\Big)\mathbf{k}\bigg),
\\
&R_3(\mathbf{k})&=
\begin{pmatrix}
0&0&1&0&0
\\
1&0&0&0&0
\\
0&1&0&0&0
\\
0&0&0&e^{-i\mathbf{k}\cdot\mathbf{a_2}}&0
\\
0&0&0&0&1
\end{pmatrix},
\end{eqnarray}
with $R(\varphi)\mathbf{k}=(\cos(\varphi) k_x-\sin(\varphi) k_y,\sin(\varphi) k_x+\cos(\varphi) k_y)$, from where the polarization at a given energy gap can be determined through \cite{Fang2012}
\begin{equation}
e^{-i\frac{2\pi}{e}p}=\prod\limits_{n\in ``occ"}\frac{\theta_n(K)}{\theta_n(K^\prime)},
\label{eq:polarizationc3}
\end{equation}
where $p=p_1=p_2$ as mentioned above, $\theta(\mathbf{k})=\bra{u_n(\mathbf{k})}R_3\ket{u_n(\mathbf{k})}$ is the expectation value of $R_3$ on band $n$ and the high-symmetry points are given by $K=(\frac{4\pi}{3},0)$ and $K^\prime=(\frac{2\pi}{3},\frac{2\pi}{\sqrt{3}})$.
\begin{figure}[ht]
	\begin{centering}
		\includegraphics[width=0.4 \textwidth,height=5cm]{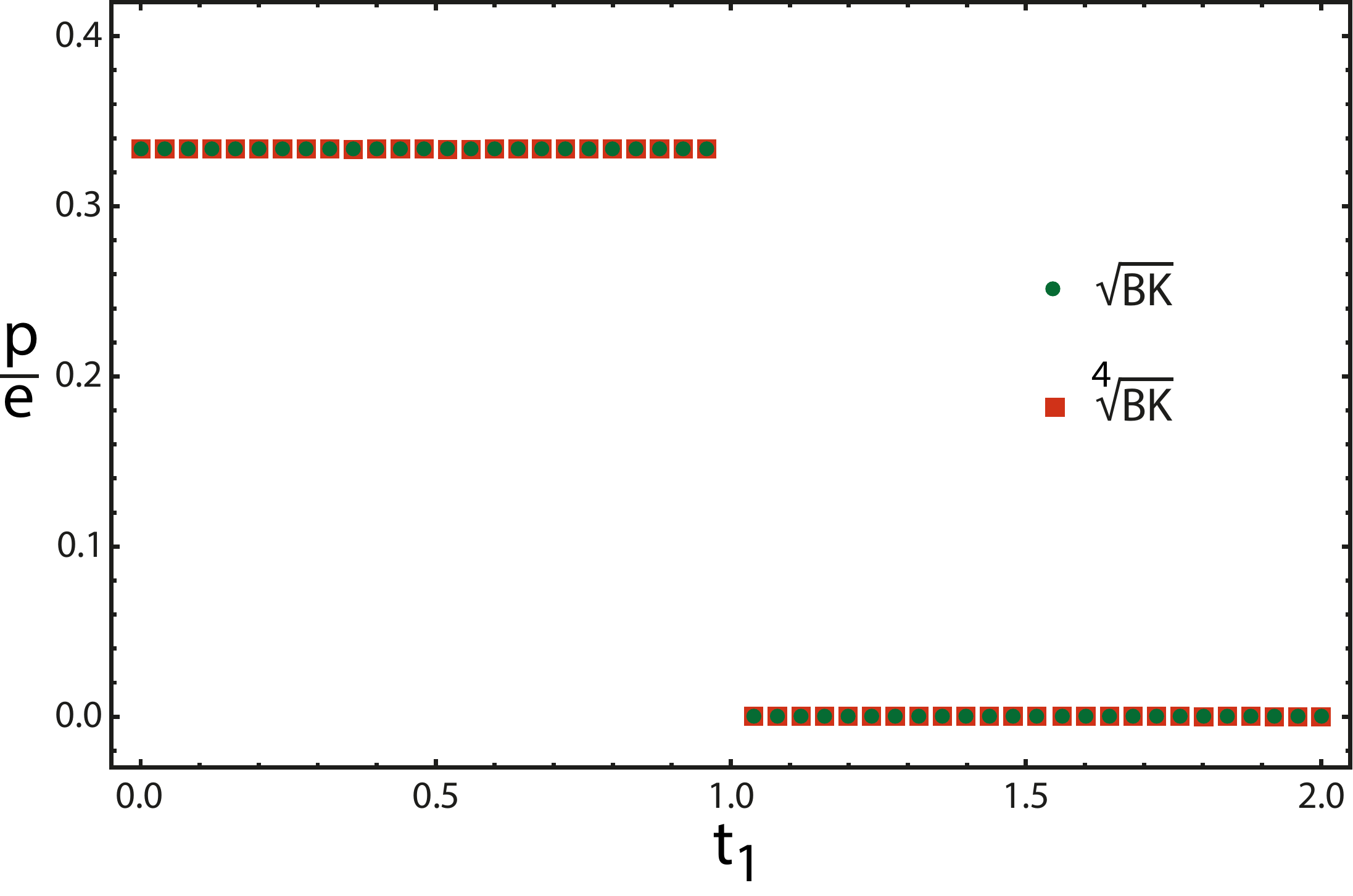} 
		\par\end{centering}
	\caption{Polarization profile, in units of $e$, along both the $\mathbf{a_1}$ and $\mathbf{a_2}$ directions for the lowest bulk band of the $\sqrt{\text{BK}}$ model (green dots) and of the $\sqrt[4]{\text{BK}}$ model (red squares).}
	\label{fig:polarizationbk}
\end{figure}
The polarization profile as a function of $t_1$, considering that only the lowest energy band is filled, is represented by the green dots in Fig.~\ref{fig:polarizationbk}.
For $t_1<1$ the polarization is quantized to $p=\frac{e}{3}$, at $t_1=t_2$ we have the gap closing point and for $t_1>1$ the polarization is trivial. 

As schematized in Fig.~\ref{fig:srkagome}(a), the $\sqrt{\text{BK}}$ model, when squared and downshifted by $c_1=t_1+t_2$ (squared on-site energy at the blue sublattice), decouples into the BK lattice and the honeycomb lattice with alternating on-site potentials, which are perturbed along the edges, with an on-site energy offset of $-t_2$, and at the corner sites, with an on-site energy offset of $-2t_2$, in both cases relative to the bulk red sites.
These impurities are the origin of the corner and edge modes of this residual lattice, which are degenerate with the respective ones from the BK lattice.
The energy spectrum of the $\sqrt{\text{BK}}$ model, as a function of a linear variation of $\sqrt[4]{t_1}$, has the form of Fig.~\ref{fig:espectrumbk4}(b), where it can be seen that in the $t_1<1$ region there are two three-fold degenerate bands of corner-states with finite and symmetric energies.
\

\subsection{$\sqrt[4]{\text{BK}}$ model}

\begin{figure*}[ht]
	\begin{centering}
		\includegraphics[width=0.85 \textwidth,height=8.25cm]{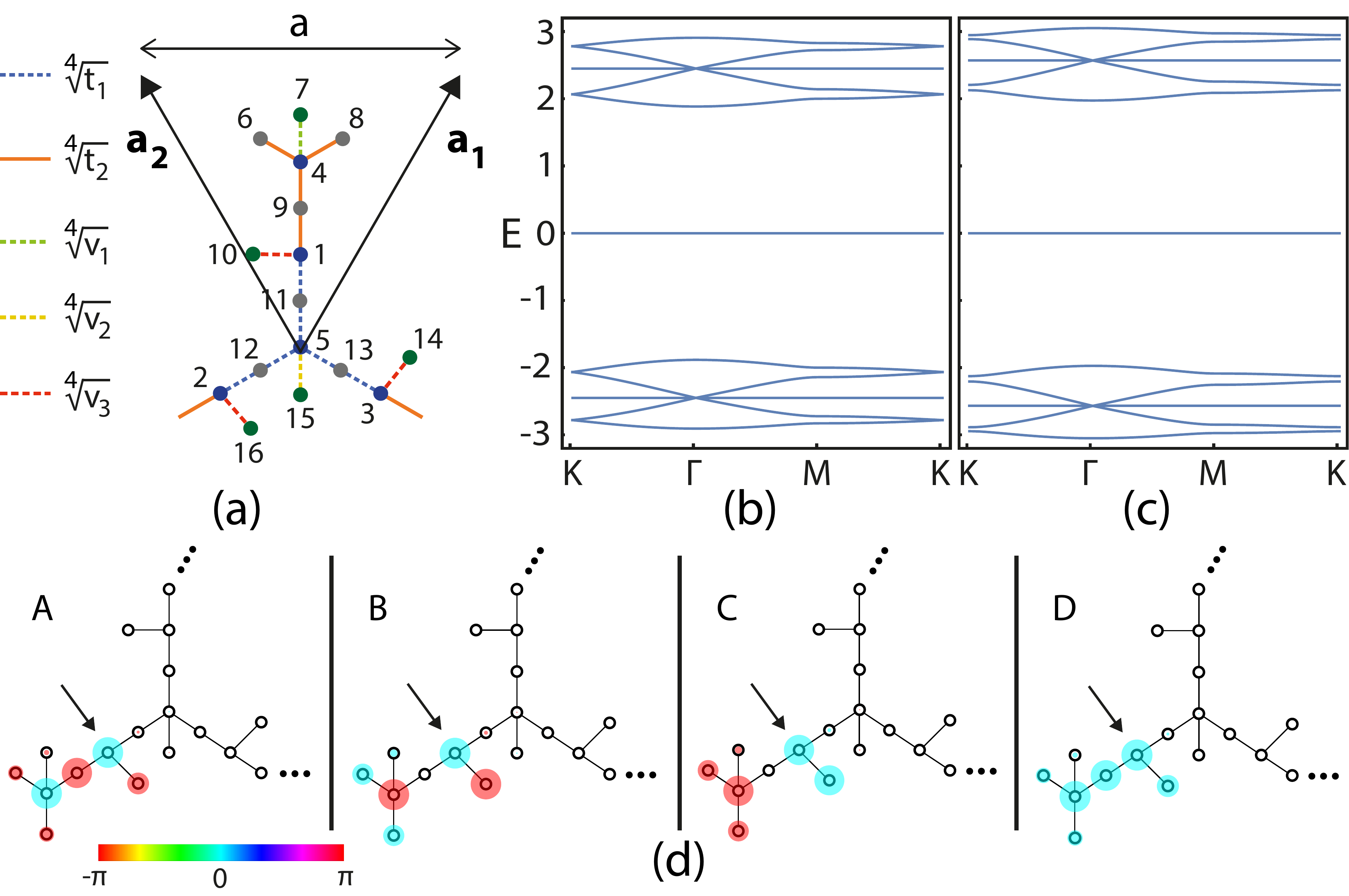} 
		\par\end{centering}
	\caption{(a) Unit cell of the $\sqrt[4]{\text{BK}}$. Blue sites form one sublattice and the green and gray sites form another.
	(b)-(c) Energy spectrum of the bulk $\sqrt[4]{\text{BK}}$ model, in units of $\sqrt{t_1}=1$, along the high-symmetry lines delimiting the orange region of the Brillouin zone depicted in Fig.~\ref{fig:srkagome}, for the same parameters as the corresponding cases there. The four gaps closings at the Dirac $K$-points in (b) are opened in (c).
    (d) Profile of the lower left corner states highlighted in Fig.~\ref{fig:espectrumbk4}(a) of the $\sqrt[4]{\text{BK}}$ triangular lattice, where the radius of the circle represents the amplitude of the wavefunction at the respective site and the color represents its phase, coded by the color bar below.
    Each state has two others degenerate in energy lying at the top and lower-right corners.
    The arrows indicate the site with the same component in all four states and corresponds, after two squaring and energy downshifting operations, to the lower-left corner site of the BK lattice depicted in the middle of Fig.~\ref{fig:srkagome}(a).}
	\label{fig:highsymbk4}
\end{figure*}
The application of the steps for the construction of the $\sqrt[4]{\text{BK}}$ model from the $\sqrt{\text{BK}}$ model leads to a lattice with the 16-sites unit cell of Fig.~\ref{fig:highsymbk4}(a).
Note that five extra green sites were included, connected to blue sites by the newly defined hopping parameters, $\sqrt[4]{v_{1(2)}}=\sqrt[4]{9t_{1(2)}}$ and $\sqrt[4]{v_3}=\sqrt{2(\sqrt{t_1}+\sqrt{t_2})}$, in order to ensure a constant squared on-site energy $c_2=3(\sqrt{t_1}+\sqrt{t_2})$ in all sites of the blue sublattice.

The bulk energy spectrum along the high-symmetry lines is shown in Fig.~\ref{fig:highsymbk4}(b) for $\sqrt[4]{t_1}=\sqrt[4]{t_2}$, where a semimetallic behavior is found when the Fermi level is placed at one of the four band touchings occurring at the $K$-point, and in Fig.~\ref{fig:srkagome}(c) for $\sqrt[4]{t_1}\neq\sqrt[4]{t_2}$, where the degeneracies are lifted at this point and four extra energy gaps appear.
In both cases asymmetric spin-1 Dirac cones centered at the $\Gamma$-point and $E=\pm\sqrt{c_2}$ appear in the spectrum.
Apart from the six-fold degenerate flat band coming from sublattice imbalance, the squared spectrum of Figs.~\ref{fig:highsymbk4}(b) and (c), downshifted by $c_2$, yields the spectra in Figs.~\ref{fig:srkagome}(b) and (c), respectively, with a global double degeneracy due to the diagonalization of $H_{\sqrt{\text{res},2^\prime}}(\mathbf{k})$, the bulk Hamiltonian of the square-root residual block.

The $\sqrt{\text{BK}}$ model has a three-fold rotation symmetry about the $z$ axis defined, in the ordered $\{\ket{j,\mathbf{k}}\}$ basis, with $j=1,2,\dots,16$, as
\setcounter{MaxMatrixCols}{20}
\begin{widetext}
	\begin{eqnarray}
	C_3:\ &R_3(\mathbf{k})&H_{\sqrt[4]{\text{BK}}}(\mathbf{k})R_3^{-1}(\mathbf{k})=H_{\sqrt[4]{\text{BK}}}\bigg(R\Big(\frac{2\pi}{3}\Big)\mathbf{k}\bigg),
	\\
	&R_3(\mathbf{k})&=
	\begin{pmatrix}
	0&0&1&0&0&0&0&0&0&0&0&0&0&0&0&0
	\\
	1&0&0&0&0&0&0&0&0&0&0&0&0&0&0&0
    \\
   	0&1&0&0&0&0&0&0&0&0&0&0&0&0&0&0
    \\
   	0&0&0&e^{-i\mathbf{k}\cdot\mathbf{a_2}}&0&0&0&0&0&0&0&0&0&0&0&0
    \\
   	0&0&0&0&1&0&0&0&0&0&0&0&0&0&0&0
    \\
   	0&0&0&0&0&0&0&e^{-i\mathbf{k}\cdot\mathbf{a_2}}&0&0&0&0&0&0&0&0
    \\
   	0&0&0&0&0&0&e^{-i\mathbf{k}\cdot\mathbf{a_2}}&0&0&0&0&0&0&0&0&0
    \\
   	0&0&0&0&0&0&0&0&e^{-i\mathbf{k}\cdot\mathbf{a_2}}&0&0&0&0&0&0&0
    \\
   	0&0&0&0&0&e^{-i\mathbf{k}\cdot\mathbf{a_2}}&0&0&0&0&0&0&0&0&0&0
    \\
   	0&0&0&0&0&0&0&0&0&0&0&0&0&1&0&0
    \\
   	0&0&0&0&0&0&0&0&0&0&0&0&1&0&0&0
    \\
   	0&0&0&0&0&0&0&0&0&0&1&0&0&0&0&0
    \\
   	0&0&0&0&0&0&0&0&0&0&0&1&0&0&0&0
    \\
   	0&0&0&0&0&0&0&0&0&0&0&0&0&0&0&1
    \\
   	0&0&0&0&0&0&0&0&0&0&0&0&0&0&1&0
    \\
   	0&0&0&0&0&0&0&0&0&1&0&0&0&0&0&0	
	\end{pmatrix},
	\end{eqnarray}
\end{widetext}
where sites 7 and 15 are assumed here collinear in the $z$-direction with sites 4 and 5, respectively, within the unit cell of Fig.~\ref{fig:highsymbk4}(a).
The polarization profile as a function of $t_1$, computed through (\ref{eq:polarizationc3}) considering that only the lowest energy band is occupied, is represented by the red squares in Fig.~\ref{fig:polarizationbk}, where it can be seen that it follows the same behavior as that of the lowest band of the $\sqrt{\text{BK}}$ model.

The energy spectrum as a function of $\sqrt[4]{t_1}$ of the $\sqrt[4]{\text{BK}}$ triangular lattice with the unit cell of Fig.~\ref{fig:highsymbk4}(a), obtained from the diagonalization of the Hamiltonian $H_{\sqrt[4]{\text{BK}}}$ with a maximum of 10 unit cells aligned along the base, is shown in Fig.~\ref{fig:espectrumbk4}(a).
Four three-fold degenerate dashed black corner bands are found, along with four continua of red edge bands appearing at different gaps, with four more continua buried within the blue bulk continua centered around $E=\pm\sqrt{c_2}$ (e.g., centered at $E=\pm\sqrt{3}\approx 1.72$ at the atomic limit $\sqrt[4]{t_1}=0$).
The zero-energy states come from the six-fold degenerate bulk band that appears due to sublattice imbalance.

The energy spectrum obtained from diagonalization of $H^{2^\prime}_{\sqrt[4]{\text{BK}}}=H^2_{\sqrt[4]{\text{BK}}}-c_2I=\text{diag}(H_{\sqrt{\text{BK}}},H_{\sqrt{\text{Res},2^\prime}})$, with $I$ the identity matrix of $\dim(I)=\dim (H_{\sqrt[4]{\text{BK}}})$ and $H_{\sqrt{\text{Res},2^\prime}}$ the square-root of the residual lattice built on the gray plus green sublattice of Fig.~\ref{fig:highsymbk4}(a), is shown in Fig.~\ref{fig:espectrumbk4}(b).
The global spectrum is a superposition of two identical spectra, one corresponding to each of the diagonal blocks, such that, e.g., each of the two dashed black curves of corner states is now six-fold degenerate.
The zero-energy states in Fig.~\ref{fig:espectrumbk4}(a) become the $E=-c_2$ states of $H_{\sqrt{\text{Res},2^\prime}}$, which fall below the energy range of Fig.~\ref{fig:espectrumbk4}(b), except for a small segment near the lower left corner for $\sqrt[4]{t_1}\lesssim 0.15$.

The energy spectrum obtained from diagonalization of
\begin{eqnarray}
H^{4^\prime}_{\sqrt[4]{\text{BK}}}&=&H^{2^\prime}_{\sqrt[4]{\text{BK}}}H^{2^\prime}_{\sqrt[4]{\text{BK}}}-c_1I \nonumber
\\
&=&
\begin{pmatrix}
H_{\text{BK}}&&
\\
&H_{\text{Res}^\prime}&
\\
&&H_{\text{Res},2^\prime}
\end{pmatrix}
\end{eqnarray} 
is shown in Fig.~\ref{fig:espectrumci4}(f), apart from the extra states with energy $E=-c_1$ ($E=c_2^2-c_1$) which fall below (above) the energy range.
The global visible spectrum consists of a superposition of four identical spectra, one coming from $H_{\text{BK}}$ [the middle lattice in Fig.~\ref{fig:highsymbk4}(a)], one from $H_{\text{Res}^\prime}$ [the right lattice in Fig.~\ref{fig:highsymbk4}(a)] and two from $H_{\text{Res},2^\prime}$.

The four labeled lower-left corner states along the $\sqrt[4]{t_1}$ line in Fig.~\ref{fig:espectrumbk4}(a) are represented in Fig.~\ref{fig:highsymbk4}(d).
For each state, two degenerate states live on the top and lower-right corners of the triangular lattice.
The site indicated by the arrows corresponds, after two squaring and energy downshifting operations, to the lower-left corner site of the BK lattice depicted in the middle of Fig.~\ref{fig:srkagome}(a). The component at this site for all four states represents a quarter of the weight of the total state, with neglible contributions from other sites of the BK lattice since we are near the atomic limit with $t_1=0.2^4\approx 0$.
As such, each corner state in Fig.~\ref{fig:highsymbk4}(d) can be written as a linear combination of the corresponding four degenerate zero-energy states at the same corner in Fig.~\ref{fig:espectrumbk4}(c) and shown to have exactly one quarter of its weight on the BK state of this set, defining the model with the unit cell of Fig.~\ref{fig:highsymbk4}(a) as the quartic-root version of the BK lattice.

\section{Quartic-root topological semimetal}
\label{sec:sr4ts}

\begin{figure*}[ht]
	\begin{centering}
		\includegraphics[width=0.95 \textwidth,height=5.cm]{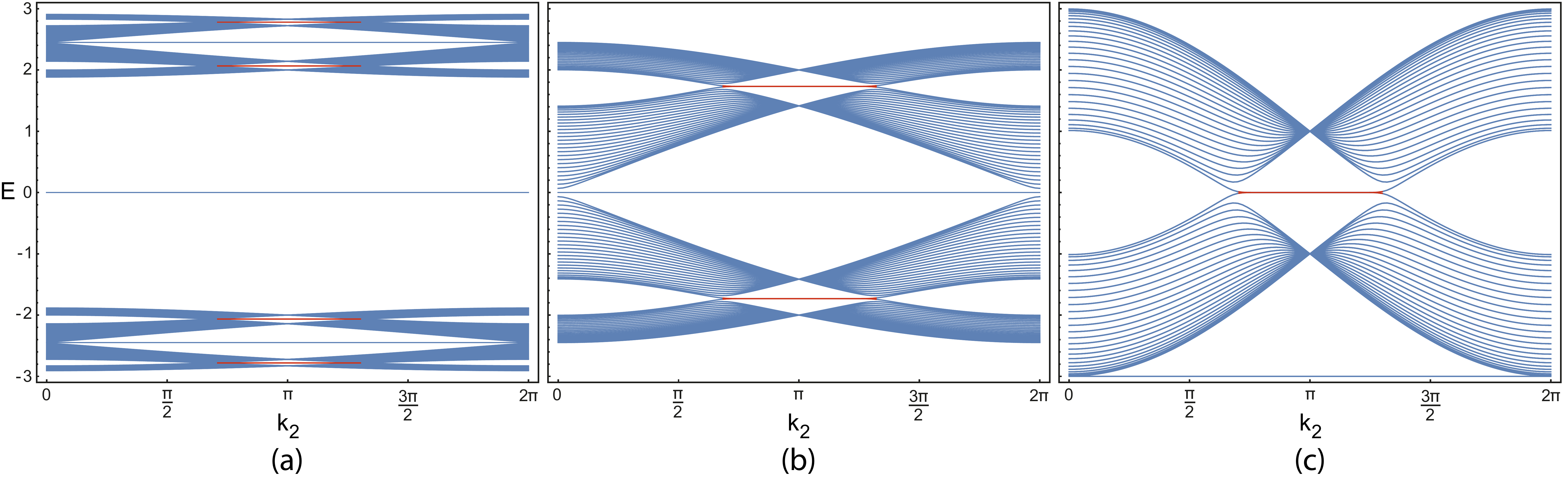} 
		\par\end{centering}
	\caption{Energy spectrum, in units of $t$, as a function of $k_2$, obtained from diagonalization of: (d) $H_{\sqrt[4]{\text{TS}}}$, the Hamiltonian of the $\sqrt[4]{\text{TS}}$ lattice with the unit cell of Fig.~\ref{fig:highsymbk4}(a) with $t_1=t_2=t$, open along $\mathbf{a_1}$ and periodic along $\mathbf{a_2}$; (b) $H_{\sqrt[4]{\text{TS}}}^{2^\prime}=H_{\sqrt[4]{\text{TS}}}^{2}-c_2I$, with $c_2=6\sqrt{t}$; (f) $H_{\sqrt[4]{\text{TS}}}^{4^\prime}=H_{\sqrt[4]{\text{TS}}}^{2^\prime}H_{\sqrt[4]{\text{TS}}}^{2^\prime}-c_1I$, with $c_1=3t$.
	Blue (red) bands correspond to bulk (edge) bands.}
	\label{fig:espectrumts4}
\end{figure*}
In this section, we expand $2^n$-root topology to semimetallic systems, using the example of the graphene ribbon as the original topological semimetal (TS) from which we construct its higher root-degree versions.
We will follow the method outlined at the end of Section~\ref{sec:qrasym2dssh}, with the change in notation HOTI$\to$TS.

\subsection{Original TS model}

We will take our original TS to be the honeycomb lattice with a uniform hopping constant $t=1$ as the energy unit for this whole section.
Considering a cylindrical square-shaped geometry with zigzag edges, where $\mathbf{a_{1(2)}}=a_{1(2)}\hat{e}_{x(y)}$ and $a_1=a_2=1$, 
with an open system along $\mathbf{a_1}$ and periodic along $\mathbf{a_2}$, allowing us to define the momentum $k_2$ along the latter direction as a good quantum number, the energy spectrum our original TS has the form of Fig.~\ref{fig:espectrumts4}(c), apart from the lowest energy flat band states, whose origin will become clear in the next sections.
The most prominent feature of this spectrum is the appearance of two degenerate flat red bands of topological zero-energy states connecting the Dirac points.

The appropriate topological invariant for this model is given by the Zak's phase along $\mathbf{a_1}$ (equivalent to the $x$-direction) of the lower band, computed at each $k_2$-value \cite{Delplace2011},
\begin{equation}
\gamma_x(k_2)=i\oint dk_1\bra{u_1(\mathbf{k})}\frac{d}{dk_1}\ket{u_1(\mathbf{k})}\mod 2\pi,
\label{eq:zakk2}
\end{equation}
where $\ket{u_1(\mathbf{k})}$ is the bulk eigenstate of the lower band.
The result is given by the solid blue dots in Fig.~\ref{fig:zakphasets}, where a non-trivial $\pi$-value is obtained for the same $k_2$-region  where the topological state are seen to lie in Fig.~\ref{fig:espectrumts4}(c).

\subsection{$\sqrt{\text{TS}}$ model}

The $\sqrt{\text{TS}}$ corresponds to the $\sqrt{\text{BK}}$ model with the unit cell shown at the lower right of Fig.~\ref{fig:srkagome}(a) in the $\sqrt{t_1}=\sqrt{t_2}=\sqrt{t}$ regime \cite{Mizoguchi2020b} since, as shown in Fig.~\ref{fig:srkagome}(b), the bulk spectrum along the high symmetry lines has semimetallic behavior when the Fermi level is set at $E=\pm\sqrt{3}$.
Notice that the roles of the squared blocks are interchanged, which is a reflection of the interchanged roles of the sublattices of the $\sqrt{\text{TS}}$, that is, gray sites 4 and 5 form now the relevant sublattice from which the the honeycomb lattice is obtained upon squaring [$H_{\text{res}^\prime}(\mathbf{k})\to H_{\text{TS}}(\mathbf{k})$], while blue sites 1, 2 and 3 form the other sublattice from which the residual lattice (the kagome lattice) is obtained upon squaring [$H_{\text{BK}}(\mathbf{k})\to H_{\text{res}^\prime}(\mathbf{k})$].
At the same time, it is now the constant squared on-site energy at the gray sublattice that defines the new energy shift, transformed according to $c_1=t_1+t_2\to c_1=3t$ (each gray site has a coordination number of three).
The squared bulk Hamiltonian of the $\sqrt{\text{TS}}$ becomes then
\begin{equation}
H_{\sqrt{\text{TS}}}^2(\mathbf{k})=c_1I_{5}+
\begin{pmatrix}
H_{\text{TS}}(\mathbf{k})&0
\\
0&H_{\text{res}^\prime}(\mathbf{k})
\end{pmatrix}.
\label{eq:hamiltts2}
\end{equation}

\begin{figure}[ht]
	\begin{centering}
		\includegraphics[width=0.4 \textwidth,height=5cm]{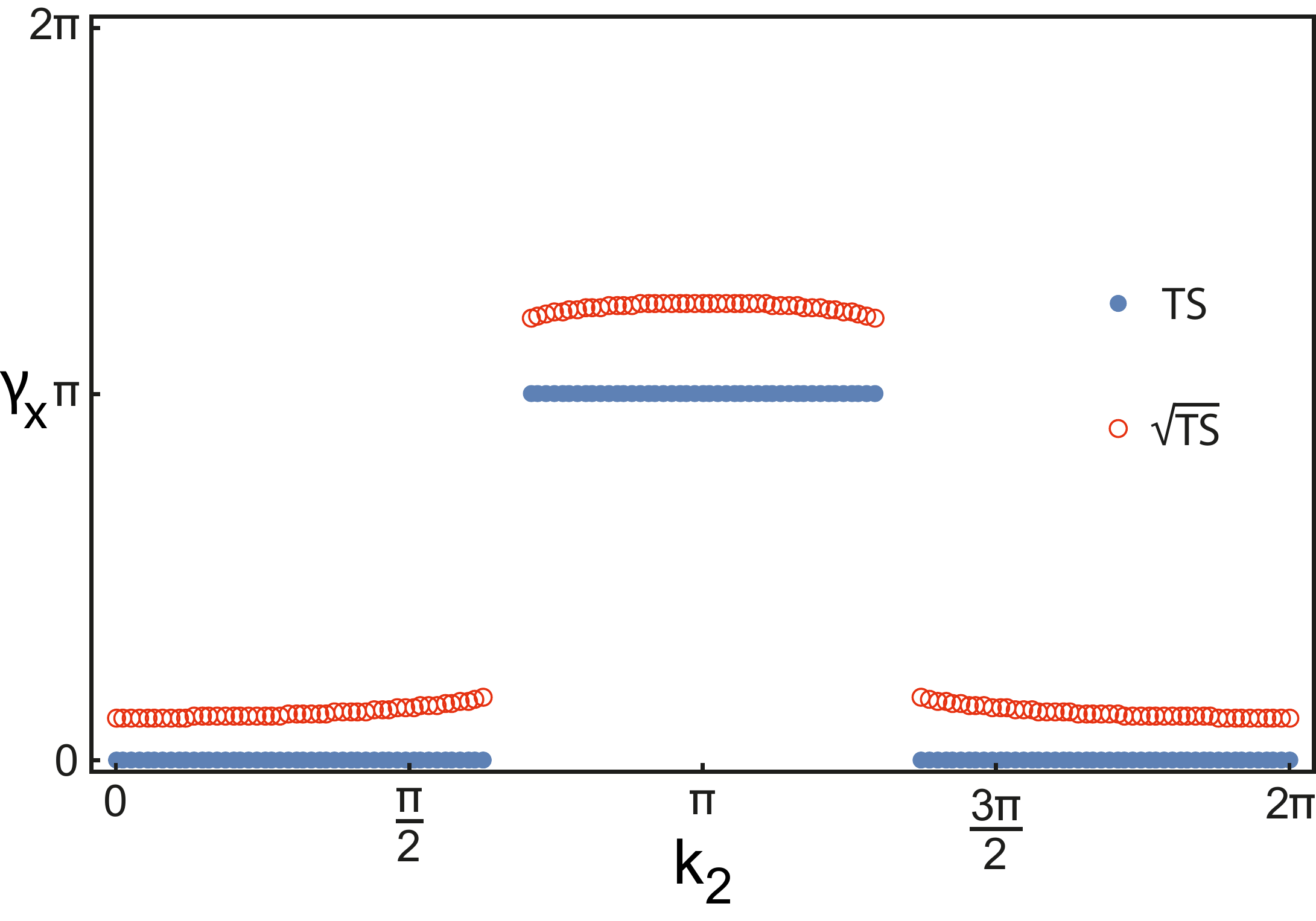} 
		\par\end{centering}
	\caption{Zak's phase along $\mathbf{a_1}= \hat{e}_x$ as a function of $k_2$ for the lowest bulk band of the TS model (solid blue dots) and of the $\sqrt{\text{TS}}$ model (open red dots).}
	\label{fig:zakphasets}
\end{figure}

Considering the $\sqrt{\text{TS}}$ in the same cylindrical geometry as in the previous subsection, its energy spectrum has the form of Fig.~\ref{fig:espectrumts4}(b). Two two-fold degenerate flat bands of edge states with symmetric finite energies are found, connecting the pairs of Dirac points appearing at $E=\pm\sqrt{3}=\pm\sqrt{c_1}$.
However, if we compute the Zak's phase of the lower bulk band along $x$ as a function of $k_2$, through (\ref{eq:zakk2}), we obtain the curves of open red points in Fig.~\ref{fig:zakphasets}, which are not $\pi$-quantized neither in the trivial nor in the non-trivial $k_2$-regions.
This model (the super-honeycomb with uniform hoppings) is thus defined as a $\sqrt{\text{TS}}$, since a quantized topological index is only recovered at the level of the topological block of the squared model, \textit{i.e.}, $H_{\text{TS}}(\mathbf{k})$ in (\ref{eq:hamiltts2}).

\subsection{$\sqrt[4]{\text{TS}}$ model}

The unit cell of the $\sqrt[4]{\text{TS}}$ model corresponds to the one depicted in Fig.~\ref{fig:highsymbk4}(a) with $t_1=t_2=t$.
As shown in Fig.~\ref{fig:highsymbk4}(b), the bulk spectrum along the high symmetry lines has semimetallic behavior when the Fermi level is set at either of the four energies where a band touching at the $K$-point occurs.
The squared bulk Hamiltonian of the $\sqrt[4]{\text{TS}}$ is written as
\begin{equation}
H_{\sqrt[4]{\text{TS}}}^2(\mathbf{k})=c_2I_{16}+
\begin{pmatrix}
H_{\sqrt{\text{TS}}}(\mathbf{k})&0
\\
0&H_{\sqrt{\text{res},2^\prime}}(\mathbf{k})
\end{pmatrix},
\label{eq:hamiltts4}
\end{equation}
with $c_2=6\sqrt{t}$ \footnote{In order to avoid extra definitions, we derived the unit cell of the $\sqrt[4]{\text{TS}}$ as that of Fig.~\ref{fig:highsymbk4}(a) in the $t_1=t_2=t$ limit, which leads to a squared on-site energy shift of $c_2=6\sqrt{t}$. However, the model could be simplified by suppressing sites 7 and 15 (along with their hopping connections $\sqrt[4]{v_1}$ and $\sqrt[4]{v_2}$) and redefining $\sqrt[4]{v_3}=\sqrt[4]{t}$. In this case, the only changes are trivial and consist of having $c_2=3\sqrt{t}$ and two fewer zero-energy flat bands in the bulk spectrum due to a lower imbalance between the residual and relevant sublattices, as sites 7 and 15 are removed from the former.}.

Considering the $\sqrt[4]{\text{TS}}$ in the same cylindrical geometry as in the previous subsections, its energy spectrum is shown in Fig.~\ref{fig:espectrumts4}(a). Four two-fold degenerate flat bands of edge states with symmetric finite energies are found, connecting the pairs of Dirac points appearing at $E=\pm\sqrt{c_2\pm\sqrt{c_1}}$.
The energy spectrum obtained from diagonalization of $H^{2^\prime}_{\sqrt[4]{\text{TS}}}=H^2_{\sqrt[4]{\text{TS}}}-c_2I=\text{diag}(H_{\sqrt{\text{TS}}},H_{\sqrt{\text{Res},2^\prime}})$, with $I$ the identity matrix of $\dim(I)=\dim (H_{\sqrt[4]{\text{TS}}})$ and $H_{\sqrt{\text{Res},2^\prime}}$ the square-root of the residual lattice, is shown in Fig.~\ref{fig:espectrumts4}(b).
The global spectrum is a superposition of two identical spectra, one corresponding to each of the diagonal blocks, such that, e.g., each topological flat band is now four-fold degenerate.
The zero-energy states in Fig.~\ref{fig:espectrumts4}(a) become the $E=-c_2$ states of $H_{\sqrt{\text{Res},2^\prime}}$, which fall below the energy range of Fig.~\ref{fig:espectrumts4}(b).

The energy spectrum obtained from diagonalization of
\begin{eqnarray}
H^{4^\prime}_{\sqrt[4]{\text{TS}}}&=&H^{2^\prime}_{\sqrt[4]{\text{TS}}}H^{2^\prime}_{\sqrt[4]{\text{TS}}}-c_1I \nonumber
\\
&=&
\begin{pmatrix}
H_{\text{TS}}&&
\\
&H_{\text{Res}^\prime}&
\\
&&H_{\text{Res},2^\prime}
\end{pmatrix}
\end{eqnarray} 
is shown in Fig.~\ref{fig:espectrumts4}(f), apart from the extra states with energy $E=c_2^2-c_1$ which fall above the energy range.
Apart from the states in the lowest energy flat band with $E=-c_1$, which stems from sublattice imbalance at the higher root-degree levels of the model, the global visible spectrum consists of a superposition of four identical spectra, one coming from $H_{\text{TS}}$, one from $H_{\text{Res}^\prime}$ and two from $H_{\text{Res},2^\prime}$.
There are now eight degenerate zero-energy topological flat bands, with only two of them coming from the topological block $H_{\text{TS}}$ and the other six from the two successive residual chains.

\section{Conclusions}
\label{sec:conclusions}

A systematic approach to the construction of $2^n$-root versions of several emblematic lattice models was presented here, generalizing to 2D systems the concept of $2^n$-root topology previously introduced for 1D chains \cite{Marques2021}.
Specifically, we constructed the 4-root version of: (i) the asymmetric 2D SSH model, shown to exhibit both 4-root weak and higher order topological states; (ii) the Haldane model with zero mass, behaving as a 4-root Chern insulator; (iii) the breathing kagome lattice in a triangular configuration, shown to host weak and higher-order boundary modes, extending the relevance of our method to non-topological systems also; (iv) the honeycomb lattice, which behaves as a 4-root topological semimetal in a cylindrical geometry, where several flat bands of topological states appear connecting different pairs of Dirac points.
The general method for the construction of these models was outlined at the end of Section~\ref{sec:qrasym2dssh}, such that their $2^n$-root versions can be straightforwardly found from the lower $2^{n-1}$-root version.

Due to their high degree of tunability, artificial lattices are the best candidates for the experimental realization and topological probing of the models introduced here.
The original and square-root versions of several models analyzed here have already been implemented in acoustic \cite{Xue2019,Zheng2019,Ni2019,Yan2020} and photonic \cite{Zhou2020,Li2020a} lattices, as well as in topoelectrical circuits \cite{Liu2019,Olekhno2020,Wu2020,Song2020,Yang2021}. This latter platform seems especially suited for implementing the $2^n$-root 2D topological insulators, since the hopping terms of the tight-binding models can be easily emulated by a judicious choice of coupled capacitors and inductors \cite{Lee2018}.

The extension of our work to higher-dimensional models, higher-order topological superconductors and more general non-Hermitian systems is an open subject that invites further studies.
Another extension would be to explicitly consider spinful models hosting spin-polarized edge and higher-order boundary modes \cite{Gladstone2021}. 
Although the method for constructing the higher-root degree versions of these systems would remain unchanged, the introduction of spin-orbit couplings (SOCs) would pose new and interesting questions.
Concretely, if the SOCs are understood as interlayer hoppings between layers of different spin projections, then the construction of the square-root model will entail, by virtue of subdivision of the lattice, the inclusion of new sites in the middle of each SOC, which can be viewed as a new layer of sites related to a specific spin projection (e.g., a fermionic spin-1/2 original TI with two spin layers from which a bosonic square-root spin-1 model with three layers can be constructed).
This prospect of having $\sqrt[2^n]{\text{TIs}}$ where the different root-degrees relate to different spin quantum number sectors, and in particular the possibility of formally deriving a bosonic model from a lower root-degree fermionic model, or vice-versa, is a very interesting one, and one that we are currently addressing.

\section*{Acknowledgments}
\label{sec:acknowledments}

This work was developed within the scope of the Portuguese Institute for Nanostructures, Nanomodelling and Nanofabrication (i3N) projects No.~UIDB/50025/2020 and No.~UIDP/50025/2020 and funded by FCT - Portuguese Foundation for Science and Technology through the project PTDC/FIS-MAC/29291/2017. A.M.M. acknowledges financial support from the FCT through the work Contract No.~CDL-CTTRI-147-ARH/2018.

\appendix

\section{Degenerate squared energy spectrum}
\label{app:degenspectrum}

In this appendix we want to show that $	H_{\sqrt{\text{par},2}}$ and $H_{\sqrt{\text{res},2}}$ in (\ref{eq:hamilthotibo42}) share the same finite-energy spectrum. 
A similar derivation can be found in the Appendix B of [\onlinecite{Ezawa2020}].
We start by writing the $\sqrt[4]{\text{HOTI}}$ bulk Hamiltonian in a manifestly chiral-symmetric form, by using the site ordering of the unit cell in Fig.~\ref{fig:ucellboti4}:
\begin{equation}
	H_{\sqrt[4]{\text{HOTI}}}(\mathbf{k})=
	\begin{pmatrix}
	0&h^R
	\\
	h^L&0
	\end{pmatrix},
	\label{eq:apphamilthotibo4}
\end{equation}
with $h^R$ ($h^L$) a $12\times 24$ ($24\times 12$) matrix.
Note that no Hermiticity condition, which would read as $h^{R^\dagger}\equiv h^L$, is imposed here. 
By squaring the Hamiltonian in (\ref{eq:apphamilthotibo4}), one can recast (\ref{eq:hamilthotibo42}) as
\begin{eqnarray}
H^2_{\sqrt[4]{\text{HOTI}}}(\mathbf{k})&=&
\begin{pmatrix}
H_{\sqrt{\text{par},2}}&0
\\
0&H_{\sqrt{\text{res},2}}
\end{pmatrix},
\label{eq:apphamilthotibo42}
\\
H_{\sqrt{\text{par},2}}&=&h^Rh^L,
\label{eq:apphamiltpar2}
\\
H_{\sqrt{\text{res},2}}&=&h^Lh^R.
\label{eq:apphamiltres2}
\end{eqnarray}
The Schroedinger equation applied to the square-root parent subspace is written as
\begin{equation}
	H_{\sqrt{\text{par},2}}\ket{\psi_{\sqrt{\text{par},2}}(k)}=E_{\sqrt{\text{par},2}}(k)\ket{\psi_{\sqrt{\text{par},2}}(k)},
	\label{eq:appschropar2}
\end{equation}
where $E_{\sqrt{\text{par},2}}(k)=c_2+E_{\sqrt{\text{HOTI}}}(k)\neq 0$ is assumed.
Inserting (\ref{eq:apphamiltpar2}) into (\ref{eq:appschropar2}) yields
\begin{equation}
	h^Rh^L\ket{\psi_{\sqrt{\text{par},2}}(k)}=E_{\sqrt{\text{par},2}}(k)\ket{\psi_{\sqrt{\text{par},2}}(k)}.
\end{equation}
Multiplying by $h^L$ from the left yields
\begin{equation}
		h^Lh^Rh^L\ket{\psi_{\sqrt{\text{par},2}}(k)}=E_{\sqrt{\text{par},2}}(k)h^L\ket{\psi_{\sqrt{\text{par},2}}(k)},
\end{equation}
which, after defining $h^L\ket{\psi_{\sqrt{\text{par},2}}(k)}:=\ket{\psi_{\sqrt{\text{res},2}}(k)}$ and using (\ref{eq:apphamiltres2}), leads to
\begin{equation}
H_{\sqrt{\text{res},2}}\ket{\psi_{\sqrt{\text{res},2}}(k)}=E_{\sqrt{\text{par},2}}(k)\ket{\psi_{\sqrt{\text{res},2}}(k)},
\end{equation}
thus showing that $E_{\sqrt{\text{res},2}}(k)=E_{\sqrt{\text{par},2}}(k)$ with regards to the finite energy spectrum.

\bibliography{srhoti}

\end{document}